\documentclass[preprint]{elsarticle}
\usepackage{setspace}
\usepackage{graphicx}
\usepackage{natbib}
\usepackage{amsmath}
\usepackage{shortcuts}
\usepackage{amssymb}
\usepackage{journals}

\begin{document}

\title{Finding the trigger to Iapetus' odd global albedo pattern:  Dynamics of dust from Saturn's irregular satellites}

\author{Daniel Tamayo\corref{cor1}}
\ead{dtamayo@astro.cornell.edu}
\address{Department of Astronomy, Cornell University, Ithaca, NY 14853}

\author{Joseph A.~Burns}
\ead{joseph.burns@cornell.edu}
\address{Department of Astronomy and College of Engineering, Cornell University, Ithaca, NY 14853}

\author{Douglas P.~Hamilton}
\ead{hamilton@astro.umd.edu}
\address{Department of Astronomy, University of Maryland, College Park, MD 20742}

\author{Matthew M.~Hedman}
\ead{mmhedman@astro.cornell.edu}
\address{Department of Astronomy, Cornell University, Ithaca, NY 14853}

\cortext[cor1]{Corresponding author, Tel: (607)255-1823}

\begin{abstract}
The leading face of Saturn's moon Iapetus, Cassini Regio, has an albedo only one tenth that on its trailing side.  The origin of this enigmatic dichotomy has been debated for over forty years, but with new data, a clearer picture is emerging.  Motivated by Cassini radar and imaging observations, we investigate Soter's model of dark exogenous dust striking an originally brighter Iapetus by modeling the dynamics of the dark dust from the ring of the exterior retrograde satellite Phoebe under the relevant perturbations.  In particular, we study the particles' probabilities of striking Iapetus, as well as their expected spatial distribution on the Iapetian surface.  We find that, of the long-lived particles ($\gtrsim$ 5 $\mu$m), most particle sizes ($\gtrsim$ 10 $\mu$m) are virtually certain to strike Iapetus, and their calculated distribution on the surface matches up well with Cassini Regio's extent in its longitudinal span.  The satellite's polar regions are observed to be bright, presumably because ice is deposited there.  Thus, in the latitudinal direction we estimate polar dust deposition rates to help constrain models of thermal migration invoked to explain the bright poles \citep{Spencer10}.  We also analyze dust originating from other irregular outer moons, determining that a significant fraction of that material will eventually coat Iapetus---perhaps explaining why the spectrum of Iapetus' dark material differs somewhat from that of Phoebe.  Finally we track the dust particles that do not strike Iapetus, and find that most land on Titan, with a smaller fraction hitting Hyperion.  As has been previously conjectured, such exogenous dust, coupled with Hyperion's chaotic rotation, could produce Hyperion's roughly isotropic, moderate-albedo surface.\end{abstract}


\maketitle

\newpage 

\section{Introduction}
Over three dozen dark irregular satellites have been discovered around Saturn using ground-based telescopes \citep{Gladman01, Sheppard03, Jewitt05, Sheppard06}.  Numerical simulations show that these irregular satellites must have undergone intense collisional evolution that would have generated large quantities of dark dust over the age of the solar system\citep{Nesvorny03, Turrini09}.  Indeed, \cite{Bottke10} estimate that on the order of $10^{20}$ kg of dust ($\sim$ a thousandth the mass of the Earth's moon) has been generated in the outer Saturnian system through collisional grinding of these satellites.  Furthermore, the recent discovery \citep {Verbiscer09} of a vast dust ring originating from the largest of the irregulars (Phoebe) shows that these dust-producing collisional processes are ongoing even today.

\begin{figure}
\includegraphics[width=9cm]{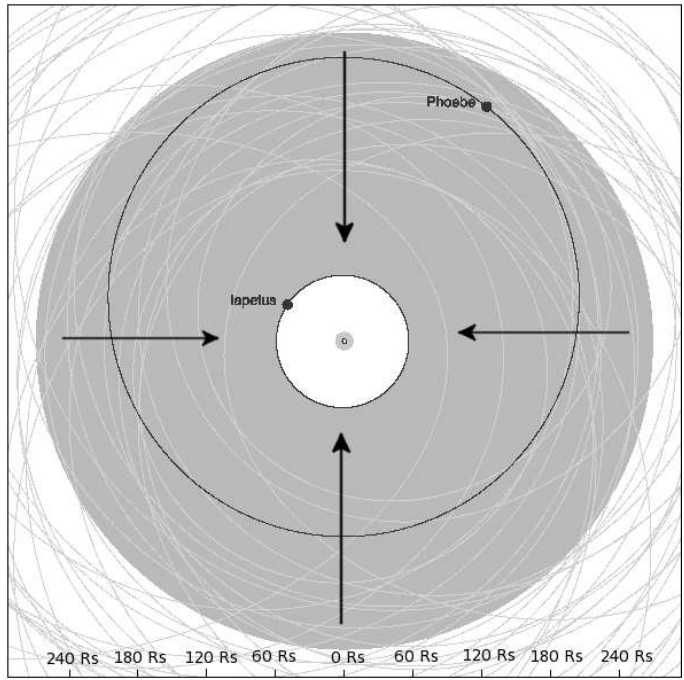}
\caption{\label{cartoon}  This schematic diagram depicts the expected extent of the Phoebe ring (solid gray), as well as the orbits of both Phoebe (black) and the rest of the irregular satellites (light gray).  The circle at the center represents the main rings, with the E ring surrounding them.  Approximate scale is provided in Saturn radii (Rs).  According to the model of \cite{Soter74}, as the ring of retrograde dust drifts inward from Phoebe's semimajor axis (arrows), Iapetus' leading side sweeps up this material, darkening its leading side.  Figure provided by Matthew S. Tiscareno.}
\end{figure}

Small dust particles are strongly affected by radiation forces \citep{Burns79}; in particular, Poynting-Robertson drag will cause particles to lose energy and slowly migrate toward their parent planet.  One should therefore expect mass transfer from the dark outer irregular satellites to the generally brighter inner regular satellites (see Fig. \ref{cartoon}).  Iapetus is the outermost of the regular satellites and, importantly, is observed to be tidally locked \citep{McCord71}.  As such, one hemisphere permanently faces the direction of motion and would plow through the cloud of dark dust as the cloud evolves inward.  Since Phoebe (and most of the other irregulars) orbits retrograde and would generate dust particles on retrograde paths, collisions with the prograde Iapetus would occur at high relative velocities ($\sim 7$ km/s) and the dust would mostly coat Iapetus only on its leading side.  This model for the exogenous origin of the dark material on Iapetus was first proposed by \cite{Soter74} and seems plausible; when one looks at the observed albedo map of Iapetus, one finds that the dark region, Cassini Regio, is centered precisely around the apex of motion---a difficult fact to explain for endogenous mechanisms \citep{Denk10}.  

As \cite{Denk10} point out, however, the extremely sharp boundaries between bright and dark material cannot be the result of simple dust deposition, which would yield more gradual transitions.  To explain the striking boundaries, as well as the bright poles, \cite{Spencer10} propose a model of runaway ice sublimation in which areas initially darkened by dust become completely blackened.  The sublimed ice then settles on the poles and on the brighter (and therefore colder) trailing side.  These two processes together, exogenous dust deposition coupled with thermal ice migration, seem the most promising mechanism to forming Iapetus' striking global albedo dichotomy.

The recently discovered "Phoebe Ring" \citep{Verbiscer09} represents a snapshot in time of the inexorable process of mass transfer from the dark outer irregulars onto the inner icy satellites.  The ring thickness implicates Phoebe as the source, showing that Soter's mechanism of coating Iapetus is ongoing.  We will show below that almost all particles in the Phoebe ring of size $\gtrsim 10 \mu$m will strike Iapetus; however, particles smaller than $\sim 5 \mu$m will strike Saturn, its main rings, or escape the system within a half-Saturn orbit($\sim 15$ yrs) due to radiation pressure \citep{Verbiscer09}.

Our presentation improves on two previously published works.  \cite{Burns96} published a short analysis of the dynamics of dust particles from Phoebe and \cite{Tosi10} included in their paper a simplified analysis that included Poynting-Robertson drag but neglected the important effects of the dominant component of solar radiation pressure (radial from the Sun), which affects particles' eccentricities and can quickly drive small grains out of the system.

This paper performs a more in-depth analysis, considering all the important radiation and tidal perturbations from the Sun and calculating the expected coverage on the Iapetus surface.  We also include the precession of Iapetus' orbital axis, which extends coverage over the poles. 

The paper is organized as follows.  Sec.2 discusses the determination of probabilities for dust striking Iapetus from numerically integrated dust orbits.  In Sec.3 we present calculated distributions of dust on the Iapetus surface, comparing them to the observed distribution and using them to obtain estimates for polar deposition rates.  Sec.4 addresses the same process for the dozens of irregular satellites other than Phoebe, and Sec.5 tracks the fate of dust grains that do not strike Iapetus and that instead collide with Hyperion and Titan.

\section{Collision Probabilities}
\subsection{Orbital Integrations for Dust Particles}
In order to estimate dust particles' likelihoods of striking Iapetus, we first consider the important effects of the perturbations affecting dust particle dynamics.  Most dust particles spend their lifetimes in a radial range (between the orbits of Phoebe and Iapetus) where the dominant perturbations are solar.  The important modifications to these particle orbits therefore result from radiation pressure and solar gravity.  Nevertheless, since small particles' eccentricities can bring them closer to Saturn, we also included the perturbation from Saturn's second-order zonal harmonic in our numerical integrations.  

As mentioned in the introduction, particles smaller than $\sim 5 \mu$m are so affected by solar radiation pressure that they are quickly removed from the system; in this size regime, electromagnetic forces from the planet's magnetosphere are negligible relative to the other perturbations \citep{Burns01}.  Note that the smallest particles might not be blown out by radiation pressure; once the particle size becomes small relative to the incident light's wavelength, the dust particles will no longer be able to effectively couple to the radiation field.  Such particles presumably account for a small fraction of the total mass, and their orbits would decay too slowly to reach Iapetus---we therefore ignore them.  

\begin{figure}[!ht]
\includegraphics[width=9cm]{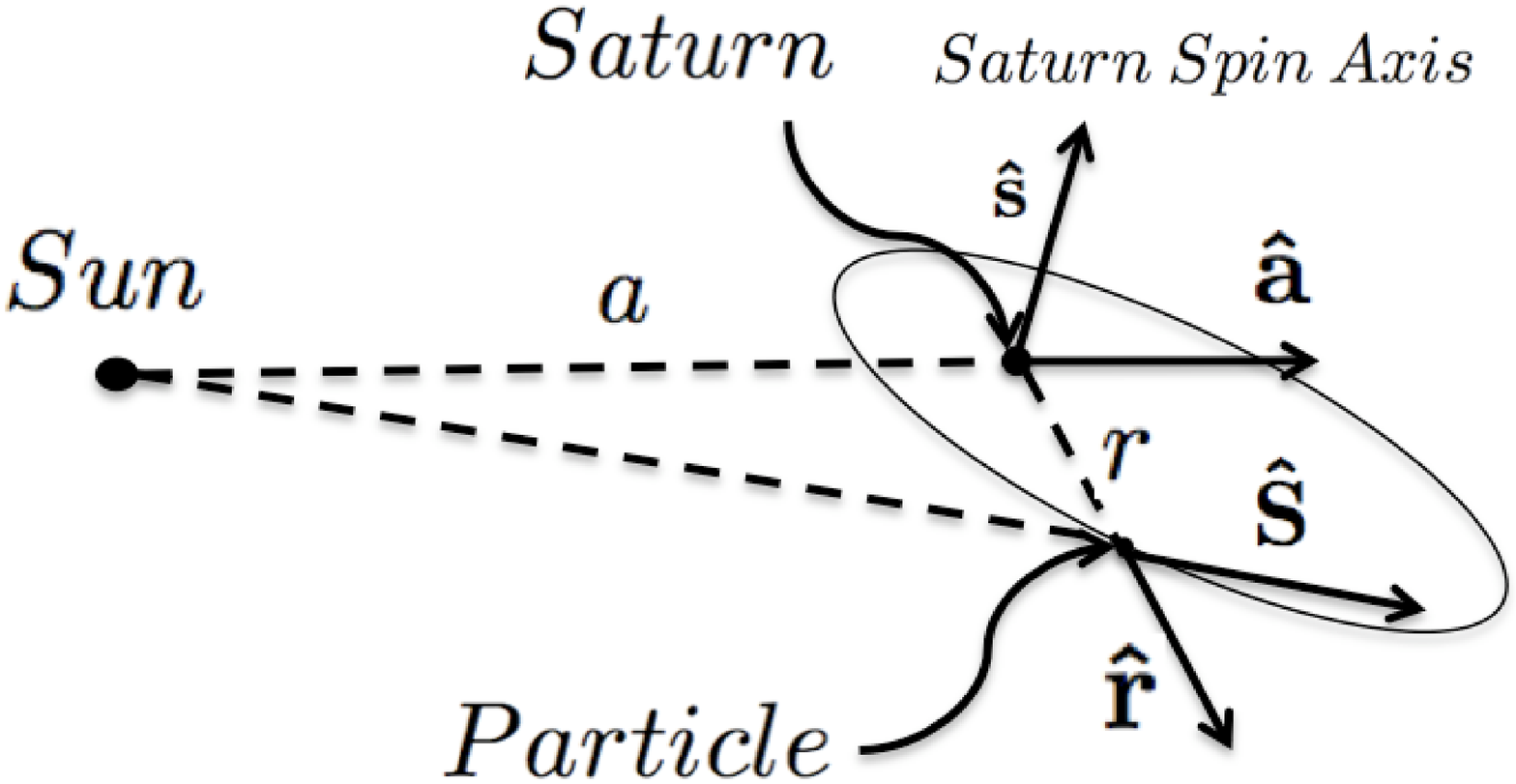}
\caption{\label{vectors} Schematic diagram showing the geometry of the important perturbations acting on dust grains in orbit around Saturn.  Vectors are described in the text above.}
\end{figure}

The equation of motion can be written as
 \begin{equation}\label{eom}
{\bf \ddot{r}} = -\frac{GM_S}{r^3} {\bf \hat{r}} + \frac{SAQ_{pr}}{mc} {\bf \hat{S}} - \frac{SA}{mc^2} Q_{pr}[({\bf \dot{r}} \cdot {\bf{\hat{S}}}){\bf \hat{S}} + {\bf \dot{r}}] - \frac{GM_{Sun}}{a^3}\nabla\Bigg({r^2 P_2({\bf \hat{a}}\cdot {\bf \hat{r}})}\Bigg) + GM_S{R_S}^2J2\nabla\Bigg(\frac{P_2({\bf \hat{s}}\cdot {\bf \hat{r}})}{r^3}\Bigg),
 \end{equation}
where the terms, in sequence, are due to the dominant Saturnian gravity, solar radiation pressure, Poynting-Robertson drag, the Sun's tidal gravity, and Saturn's J2.  $G$ is the gravitational constant, $M_S$ Saturn's mass, $r$ the dust particle's distance from Saturn, $S$ the solar flux at the particle's position, $A$ the particle's cross-sectional area, $Q_{pr}$ the grain's pressure efficiency, $m$ the particle mass, $c$ the speed of light, $a$ the semi-major axis of Saturn (assumed to be on a circular orbit about the Sun), $R_S$ the radius of Saturn, $J2$ Saturn's second-order zonal harmonic, and $P_2$ the second Legendre polynomial.  The vector ${\bf \dot{r}}$ is the particle's velocity, and the other vectors can be seen in Fig. \ref{vectors}; ${\bf \hat{r}}$ is the direction from Saturn to the particle's position, ${\bf \hat{S}}$ is the direction from the Sun to the particle position, ${\bf \hat{a}}$ is the direction from the Sun to Saturn, and ${\bf \hat{s}}$ is the direction along Saturn's spin axis (perpendicular to the equatorial plane).  

While commonly considered in a heliocentric context, Poynting-Robertson drag also causes particles' orbits around a host planet to decay into the planet on a timescale given by \citep{Burns79}:

\begin{equation}\label{pr}
\tau_{PR} = 530 \: years \times \frac{a_{Sat}^2}{\beta_{R/G}},
\end{equation}
where $a_{Sat}$ is Saturn's semimajor axis in AU ($\approx 9.5$) and $\beta_{R/G}$ is the dimensionless ratio of the radiation force to the {\it Sun's} gravitational force (in this case approximately $0.36 /r$, where {\it r} is the particle size in $\mu$m).  $\tau_{PR}$ therefore scales linearly with particle size.

Superimposed on this slow orbital decay (timescale $\sim 10^{6}$ years for spherical 10 $\mu$m particles) is a fast oscillation in the eccentricity (P $\sim$ 1 Saturn yr $\simeq$ 30 yrs) due to both solar radiation pressure and the Sun's tidal gravity \citep{Burns79, Hamilton96}.  Eventually, dust-particle orbits will cross that of Iapetus as Poynting-Robertson drag reduces the orbit size and radiation pressure periodically induces large eccentricities.  Over time, therefore, the dark particles will impact Iapetus' leading side.

As opposed to gravitational accelerations, accelerations due to radiation forces are mass---and therefore size---dependent.  As a result, we numerically integrate orbits for different-sized particles using the well-established dust integrator ``dI" \citep[see][]{Hamilton93, Hamilton96, Hamilton08}.  This provides a particle's orbital elements as a function of time for each particle size.  

In any particular history, we choose particles of a given size and assign them a density (we assume that dust particles would share Phoebe's density of 1.6 g/$cm^{3}$).  As discussed in further detail below, they are then started at various positions along Phoebe's orbit and initially move with Phoebe's velocity.  We determine that for our assumed density, particles smaller than 4 $\mu$m are so affected by radiation pressure that within the first half-Saturn year their eccentricities reach a value of unity and the grains either collide with Saturn or its rings, or escape the Saturn system entirely.  This corresponds to $\sim 10$ particle orbits and a negligible probability of collision with Iapetus.  One should therefore expect only a significant contribution to Iapetus from particles $\gtrsim$ 4 $\mu$m in size.  Since dust particles are not actually spherical and will contain some void space, our assumed density is probably high and our 4 $\mu$m likely represents a lower limit.  

\begin{figure}[!ht]
\includegraphics[width=9cm]{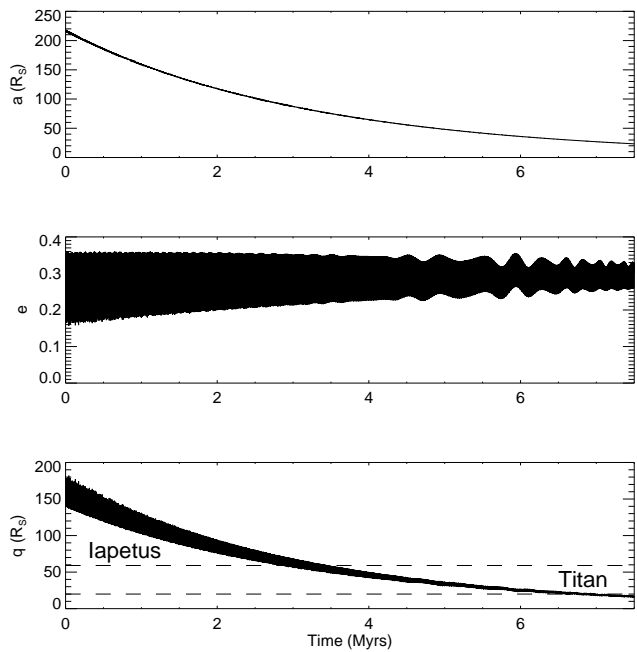}
\caption{\label{integration} Evolution of a $25 \mu$m particle under the effects of solar perturbations.  Top panel shows the particle's semimajor axis, which starts at Phoebe (a $\approx 215 R_S$) and decays on a timescale $\sim 2.5$ Myrs.  Superimposed on this slow evolution of the semimajor axis is a rapid oscillation in the eccentricity (middle panel) on a timescale $\approx 1$ Saturn year $\approx 30$ yrs.  The bottom panel shows the particle orbit's pericenter $q$, along with the semimajor axis of Iapetus and Titan.}
\end{figure}

On the other extreme, the orbital eccentricities of particles larger than 500 $\mu$m (Poynting-Robertson decay timescale $\gtrsim$ 50 million years) are almost completely unaffected by radiation forces and are dominantly affected by the Sun's tidal gravitational force, which is independent of particle size.  We therefore run integrations for particle sizes of 5, 10, 25, 50, 100, and 500  $\mu$m.  A 25 $\mu$m particle's orbital element evolution is shown in Fig. \ref{integration} with its slow semimajor axis decay and rapid eccentricity oscillations($P\sim30$ yrs).  The bottom panel shows the pericenter distance $q$.  When $q$ crosses a satellite's semimajor axis, collisions with that moon become possible.

A few considerations supply the appropriate initial conditions for the integrations.  All particles leaving Phoebe must have initial speeds $\gtrsim$ Phoebe's escape speed $v_{esc}$.  Since dust-producing impact events produce a distribution of ejecta velocities with a decaying tail toward higher speeds, one should expect most particles that escape Phoebe to have launch speeds near $v_{esc}$ \citep{Farinella93}.  Therefore, since Phoebe's escape velocity is much smaller than its orbital velocity ($\sim$ 0.1 km/s vs. $\sim$ 1.7 km/s), we expect most dust particles generated in an impact with Phoebe to approximately share that moon's orbital elements.  This sets the initial conditions for the semimajor axis, eccentricity and inclination ({\it a} = 1.296$\e{7}$ km, {\it e}  = 0.156, {\it i} = $175.2^{\circ}$ with respect to Saturn's orbital plane about the Sun).  

\begin{figure}[!ht]
\includegraphics[width=9cm]{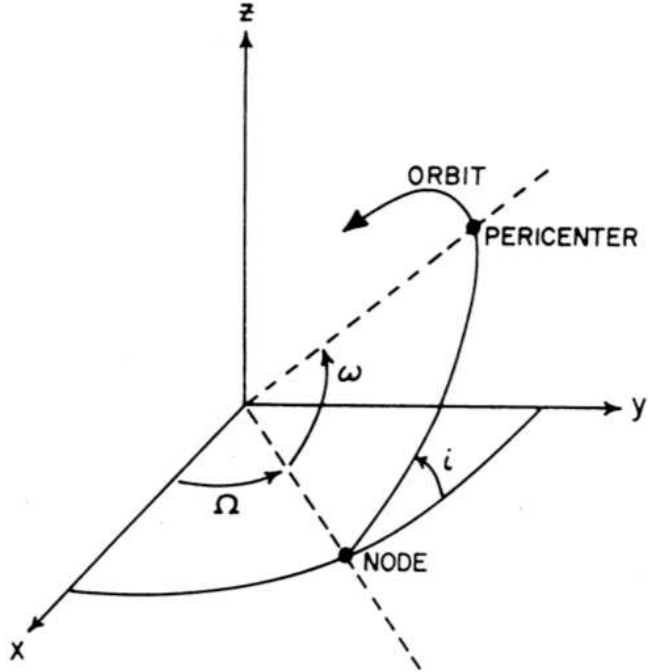}
\caption{\label{orbit} The three Euler angles that define an orbit's orientation:  $i, \Omega$ and $\omega$.  \citep[Figure from][]{Greenberg82}.}
\end{figure}

The last three initial conditions---the three angles that determine the orientation of the orbit (see Fig. \ref{orbit})--- depend on the time of impact itself.  Specifically, they are set by the orientation of Phoebe's orbit ({\it $\Omega$} and {\it $\omega$}), and Phoebe's position within its orbit ({\it f}, the true anomaly) at the time of impact.  This would represent a formidable phase space to cover for long integrations, but fortunately several considerations limit the phase space considerably.

The shortest timescale for the perturbations involved is the $\sim 30$-yr period of the Sun's apparent motion about Saturn.  Since the dust particles' orbital periods around Saturn are much shorter than this ($\approx$ 1.5 years), the exact position of Phoebe ({\it f}) in its orbit at the time of impact does little to influence the subsequent evolution of the orbit's shape or orientation; thus it can be chosen arbitrarily.

Phoebe's orbit orientation at the time of impact, however, {\it is} important because it precesses more slowly.  Nevertheless, one can still limit the phase space by inspecting the geometry.  Since the dominant perturbations in this problem are all of solar origin, the logical plane from which to reference inclinations is Saturn's orbital plane (i.e., the plane in which the Sun appears to move in a Saturnocentric frame).  Phoebe's inclination relative to Saturn's orbital plane of $175^{\circ}$ means its own orbital plane is almost coplanar, albeit in a retrograde sense, with this reference plane.

In the limit of coplanarity, only one angle (rather than both {\it $\Omega$} and {\it $\omega$}) is required to specify the orientation of the orbit given the inclination, i.e., the angle between an arbitrary reference direction and the orbit's pericenter.  In this case, the physically meaningful reference direction is the one toward the source of perturbations, the Sun.  The orbital evolution of the dust particle therefore does not depend strongly on {\it $\Omega$} and {\it $\omega$} independently, but rather on the combination {\it $\Omega$} - {\it $\omega$}, which specifies the angle from the Sun's direction to pericenter.  Note that for a prograde orbit, the angle from the Sun's direction to pericenter would be {\it $\varpi \approx \Omega + \omega$}, but since {\it $\omega$} is measured in the direction of orbital motion, the appropriate combination for {\it retrograde} orbits is {\it $\Omega$} - {\it $\omega$}.

The approximations discussed above transform an intractable multidimensional space of initial conditions into a simple one-dimensional space.  The elements {\it a}, {\it e}, and {\it i} are those of Phoebe's orbit, and the only other initial condition left to supply is the quantity $\Omega - \omega$, with $ \Omega$ measured relative to the Sun's direction at the time of impact.  Since impacts could happen at any point in the precession cycle, we chose to perform integrations for eight equally-spaced values of $\Omega - \omega$.  

We therefore generate, for each particle size, eight sets of {\it a(t)}, {\it e(t)}, and {\it i(t)} corresponding to eight equally-spaced initial values of $\Omega - \omega$.  Taking the initial values of $\Omega - \omega$ as equally likely, we average over the eight sets of outputs, yielding, for each particle size, a single set of functions {\it a(t)}, {\it e(t)}, and {\it i(t)}.  These provide the inputs for the collision probability calculations.  While we exploit Phoebe's orbit's near-alignment with Saturn's orbital plane to combine $\Omega$ and $\omega$ for our initial conditions, the numerical integrations are carried out fully in three dimensions.  This allows us to track the orbital inclination, a crucial input to a 3-D collision probability calculation.  

\subsection{Collision Probabilities}
In order to estimate collision probabilities, we used the formalism developed by \cite{Greenberg82}, as improved by \cite{Bottke93}.  In this formalism, the dust particle's and Iapetus' semimajor axes, eccentricities and inclinations ({\it a, e, i}) are taken as known while the precession angles that determine the orientation of the orbits ($\Omega$ and $\omega$) are treated as uniformly distributed.  Barring resonances between the orbital periods of the dust particles with Iapetus, this should be a good assumption over collision timescales ($\gtrsim 10^6$ years), which are long compared to the longest precession timescale (Iapetus' orbit pole, $\tau \sim 10^{3}$ years).  This assumption was found to agree with the angular distributions from the numerical integrations.  

An alternative strategy could have been to numerically integrate many dust particle orbits and to directly see when and where on Iapetus they strike.  One drawback of such a method is that Iapetus' small size relative to the dust orbits would dictate using extremely small step sizes in the integration.  Our approach allowed us to perform fewer computationally expensive orbit integrations per particle size in exchange for computationally cheaper collision probability integrals.

The calculations of \cite{Greenberg82} are too complicated to reproduce here.  The calculation is performed by first calculating the values of $(\Omega_p, \Omega_I, \omega_p, \omega_I)$ that would lead to the two orbits crossing (`p' subscripts refer to the particle and `I' subscripts to Iapetus).  Then the objects' finite size is taken into account by Taylor-expanding around these crossing solutions to find the volume in $(\Omega_p, \Omega_I, \omega_p, \omega_I)$ space over which collisions are possible.  Finally one calculates from Keplerian theory the probability that both objects will simultaneously be close enough to the point of closest approach for a collision to occur within one object's orbit.  The ratio of this probability to the orbital period provides a collision frequency.  We compared our code to the test cases presented in \cite{Bottke93} and found it reproduced their results.  

Due to the wide disparity between orbital period ($\sim$ 1 year) and the collision timescale ($\sim 10^6$ years), it is impractical to calculate collision probabilities for every orbit.  One can see in Fig. \ref{integration}, however, that while the eccentricity is oscillating rapidly, the envelope that bounds the oscillation changes slowly, on roughly the Poynting-Robertson timescale ($\tau_{PR} \gtrsim$ 1 million years).  In particular, the figure uses only $10^4$ equally-spaced points in time and is still able to capture the full behavior.  As a result, rather than calculating probabilities every orbit, we did so for $10^4$ timesteps.

Given {\it a, e, i} for both Iapetus and dust particle, Greenberg's formalism (1982) provides a collision frequency,
\begin{equation}\label{frequency}
Frequency = \frac{Probability \: of \: collision \: within \: one \: orbit}{Period \: of \: orbit}.
\end{equation}
For timesteps $\Delta t \ll$ 1 / {\it Frequency}, one can then straightforwardly express the collision probability within $\Delta t$ as
\begin{equation}\label{P}
P = Frequency \times \Delta t.
\end{equation}
One can then recursively generate a cumulative probability of collision $C$, i.e., the probability at time $t$ that the particle has already struck Iapetus.  Starting with $C$(0) = 0,
\begin{equation}\label{recursive}
C(t_i) = C(t_{i - 1}) + (1 - C(t_{i - 1}))*P(t_i).
\end{equation}
The probability of two collisions within a single $\Delta t$ is negligible and was ignored.  $P(t_i)$ depends on the orbital elements for Iapetus and the dust particle at $t_i$.  For the dust particle we used {\it a($t_i$), e($t_i$)}, and {\it i($t_i$)}, generated as described above.  For Iapetus, we used the present values of {\it a} $=$ 3.561\e{6} km and {\it e} $=$ 0.03.  Iapetus' inclination with respect to Saturn's orbital plane, however, changes significantly over time and must be considered more carefully.  

\begin{figure}[!ht]
\includegraphics[width=9cm]{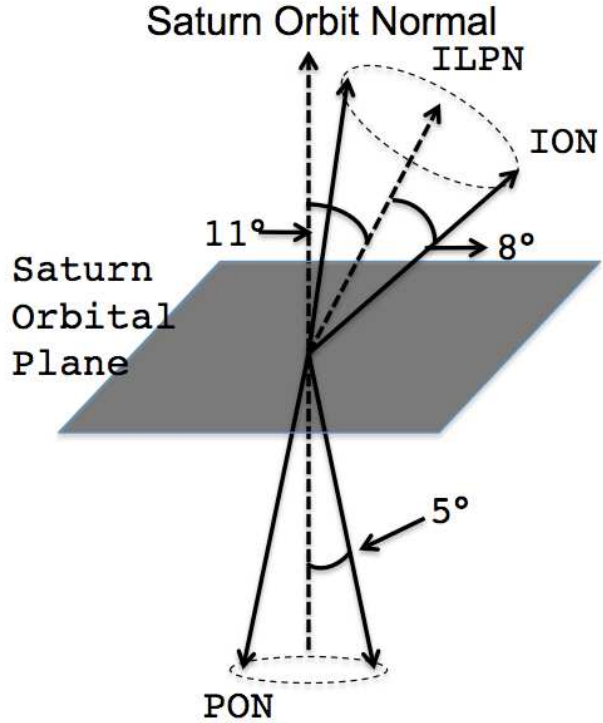}
\caption{\label{Laplace} A schematic representation of the changing orientations of Iapetus' and Phoebe's orbits (represented by their respective orbit normals PON $=$ Phoebe Orbit Normal and ION $=$ Iapetus Orbit Normal).  The moons' orbit normals precess at constant inclinations ($5^{\circ}$ and $8^{\circ}$ for Phoebe and Iapetus, respectively) to the normal vector to their local Laplace planes, sweeping out a cone.  Phoebe's Laplace plane coincides with Saturn's orbit normal, while Iapetus' local Laplace plane normal (ILPN) is inclined about $11^{\circ}$ to Saturn's orbit normal.}
\end{figure}

To first approximation, the orbit normal precesses uniformly and at a constant inclination to a vector determined by the perturbations causing the precession.  This causes the orbit normal to sweep out a cone (see Fig.\ref{Laplace}).  The vector around which orbits precess defines the local Laplace plane (normal to this vector).  At Phoebe's orbit, all the dominant perturbations are solar, so the local Laplace plane corresponds to the plane in which the Sun appears to move, Saturn's orbital plane.  Close to Saturn, where the dominant perturbation is Saturn's oblateness, the local Laplace plane is Saturn's equatorial plane.  Iapetus has the unique orbital property among satellites of existing at a distance where Saturnian and solar perturbations are comparable, and the local Laplace plane is intermediate, at about $11.5^{\circ}$ to Saturn's orbital plane \citep[see][]{Ward81}.

Because of this misalignment, although Iapetus will precess at approximately constant inclination to the normal to its local Laplace plane, its inclination relative to our reference plane (Saturn's orbital plane) will change as the orbit precesses (see Fig. \ref{Laplace}).  We therefore assumed uniform precession and coarsely averaged the probability calculation over an entire precessional cycle, sampling more finely when the inclinations of the particle and Iapetus were antiparallel and the collision probability was changing fastest.  

Finally, as mentioned in Sec.\hspace{0.25pc}2.1, we averaged over the eight equally probable initial conditions that we integrated, yielding an overall cumulative probability of collision for the given particle size.  

Apart from Iapetus, we also tracked collisions with Hyperion and Titan, as well as re-impacts into Phoebe.  We therefore straightforwardly generalized the discussion above to not only update the cumulative probability of collision with Iapetus at each timestep, but also those with the other three moons.  As is discussed below, Titan's large size renders it a sink for any long-lived dust particles that cross its path; thus, no other moons interior to it would receive appreciable amounts of dust and such bodies are therefore not tracked.  As mentioned earlier, however, a significant fraction of the particles smaller than $\sim$ 5 $\mu$m whose eccentricities all reach unity will strike Saturn or its rings within the first half Saturn-year.  

\subsection{Results}
\begin{figure}[!ht]
\includegraphics[width=9cm]{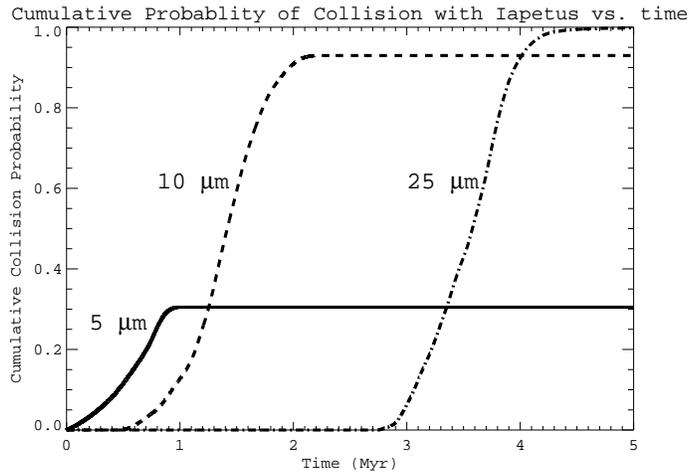}
\caption{\label{probabilities} Cumulative collision probabilities vs. time for 5, 10 and 25 $\mu$m particles.  Particles $\gtrsim 10 \mu$m almost all strike Iapetus, though larger particles take a longer time to do so.}
\end{figure}

Figure \ref{probabilities} shows the calculated cumulative collision probability with Iapetus for 5, 10, and 25 $\mu$m grains.  

Particles 10 $\mu$m and larger, being less affected by radiation forces, evolve inward via Poynting-Robertson drag so slowly (i.e., they execute many Iapetus-crossing orbits before crossing the orbits of Hyperion or Titan) that they almost all eventually strike Iapetus.  As stated before, particles smaller than about 4 $\mu$m quickly strike Saturn or escape the Saturn system.  Therefore, of the longer-lived particles, {\it almost all} particle sizes are bound for Iapetus---only a very narrow size range (between about 4 and 10 $\mu$m for the chosen density) can miss and end up mostly on Titan, with a substantially smaller fraction striking Hyperion.

We mention that, while almost all particles larger than $\sim$ 10 $\mu$m would eventually strike Iapetus, it takes larger particles longer to evolve inward and hit the satellite.  In particular, for particles in the geometrical optics limit (the peak wavelength in the solar spectrum $\sim 0.5 \mu$m $\ll 2\pi r_{dust}$, satisfied for all particle sizes we consider), the Poynting-Robertson decay timescale grows linearly with particle size \citep[see][]{Burns79}.  As a reference, assuming particles share Phoebe's density of 1.6 g/$cm^{3}$, 10 $\mu$m particles reach Iapetus in $\approx 1$ Myr.  

As particle sizes increase, one should expect to find a threshold where particles stop hitting Iapetus when the Poynting-Robertson decay timescale becomes longer than the timescale for the destruction of dust grains.  Unfortunately, destruction lifetimes for dust in the outer Saturn system are not well constrained \citep[cf.][]{Burns01}.  One mechanism for the destruction of dust grains is through mutual collisions.  One can estimate the mean free time between particle collisions as
\begin{equation}\label{mft}
t_{MF} \sim \frac{P}{\tau}
\end{equation}
where P is the particles' orbital period and $\tau$ the ring's normal optical depth.  Taking the optical depth in the Phoebe ring, $\tau \sim 2 \times 10^{-8}$ \citep{Verbiscer09}, this yields $t_{MF} \sim 100$ Myr, the Poynting-Robertson decay timescale corresponding to 1 mm grains; however, for each particle size, only collisions with particles of roughly the same size or larger affect the dynamics.  This would act to increase $t_{MF}$, but is dependent on the (currently unconstrained) particle size distribution.  On the other hand, dust rings collisionally generated early in the Solar System likely had higher optical depths \citep{Bottke10}, lowering $t_{MF}$.  For this work we chose the maximum upper-size cutoff imposed by setting the Poynting-Robertson decay timescale equal to the lifetime of the Solar System.  This yields a particle size of $\sim 1$ cm.  Improved estimates of collisional dust lifetimes in the Phoebe ring must await further observations. 

As the introduction mentions, a large supply of dust has been available in the outer Saturn system over the course of the Solar System's history.  The fact that particles $\gtrsim$ 10 $\mu$m are virtually certain to strike Iapetus strongly implicates collisionally generated dust as the trigger to Iapetus' stark albedo dichotomy.  An exogenous origin of the dark material explains why the pattern is centered on the apex of Iapetus' motion and, as shall be shown in Sec. 3, the dynamics predict a wrapping of dark material onto the trailing side consistent with that observed.  

\subsection{Titan, the gatekeeper to the inner Saturnian system}

We now explain the sharp drop in the final fraction of particles that strike Iapetus between 5 and 10 $\mu$m, as seen in Fig. \ref{probabilities}.  This is due to the large eccentricities induced by radiation pressure, visible in Fig. \ref{integration}.  For the smallest particles, the eccentricities are high enough that before the dust grains' probabilities of striking Iapetus near certainty, their orbits begin to cross that of Titan.  Saturn's largest moon is such a better interceptor of particles that the probability of striking Iapetus quickly stops increasing and levels off.  

There are several reasons why Titan is highly efficient at eliminating dust particles.  Most obviously, its sheer size makes its geometrical cross section larger than Iapetus' by a factor of about 12.  Another reason is that collision rates depend on the objects' relative velocity, as this determines how frequently the objects can potentially encounter each other (see discussion following Eq. \ref{Tcol}).  Relative velocities between dust particles and Titan are substantially higher than those with Iapetus simply because in order to reach the further-in Titan, particles generally have to be on very eccentric orbits ({\it e} = 0.7-0.9), and will encounter Titan close to periapse.

One might have expected slow relative velocities to lead to enhanced collision probabilities due to strong gravitational focusing for slow encounters.  In fact, gravitational focusing plays little role in this problem because the moon orbits are prograde while the dust orbits are retrograde, resulting in high relative velocities compared to the satellites' escape velocities ($v_{esc} = 0.572$ km/s for Iapetus, $v_{esc} = 2.639$ km/s for Titan).  For typical encounter velocities, Iapetus' gravitational cross-section is about 0.5\% greater than its geometrical cross-section ($\lesssim 10\%$ for Titan).  

Though we account for gravitational enhancements to the collision cross section, in our orbit integrations we ignore close encounters with Titan (and all other satellites) on subsequent orbital paths.  For typical relative velocities, the maximum scattering angle from a close encounter is $\approx 10^{\circ}$.  The corresponding angle for Iapetus is $\approx 0.5^{\circ}$.

We postpone our discussion of the total amount of material that strikes Titan and the smaller Hyperion, along with its implications, until Sec. 5.

\section{Coverage}
Since much of the dust previously orbiting in the outer Saturnian system will eventually strike Iapetus, we ask where on Iapetus those particles would have landed.  In particular, can the dynamics match the extent of Iapetus' dark side, Cassini Regio?  The emplacement of dust could then trigger the thermal migration of ice thought to give Iapetus the striking appearance it has today \citep{Spencer10}.

\begin{figure}[!ht]
\includegraphics[width=9cm]{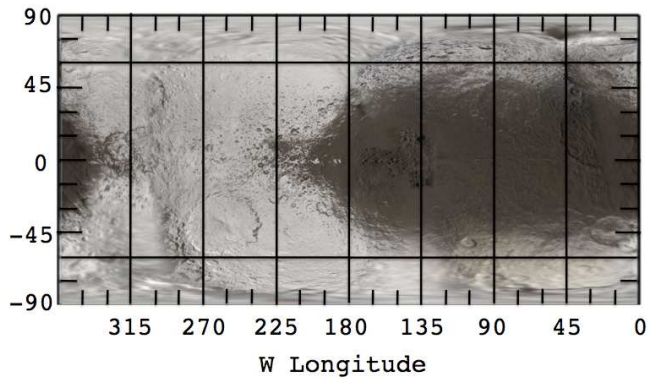}
\caption{\label{iapetusmap} Global mosaic of Iapetus \cite[from][]{Albers}.  Dark Cassini Regio is centered around the apex of Iapetus' motion, roughly at $90^{\circ}$W, and extends tens of degrees beyond $0^{\circ}$ and $180^{\circ}$W onto the trailing side.  The bright poles (beyond $\sim \pm60^{\circ}$ latitude) and sharp boundaries between light and dark terrain are likely the result of thermal ice migration \citep{Spencer10}.}
\end{figure}

Cassini Regio extends beyond Iapetus' leading side by tens of degrees onto the trailing side along the equator (see Fig. \ref{iapetusmap}).  As \cite{Burns96} have suggested, dust eccentricities naturally explain the longitudinal extension of the dark material onto the trailing side.  

If orbits were perfectly circular, dust particles would only strike Iapetus head-on, as both objects would be moving perfectly azimuthally in opposite directions; thus, only the leading face would be darkened since, as discussed at the end of Sec. 2.4, the encounter velocities make gravitational focusing negligible.  

\begin{figure}[!ht]
\includegraphics[width=9cm]{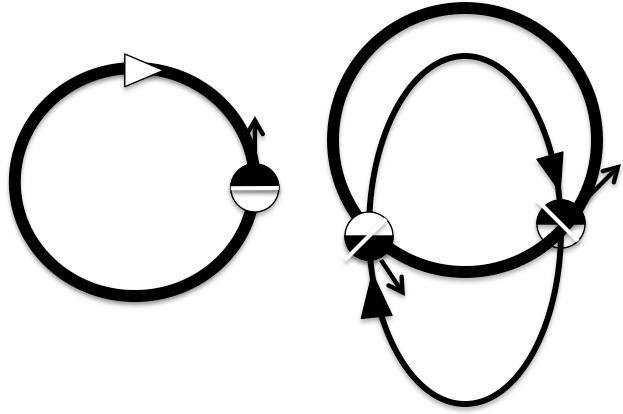}
\caption{\label{extensiondiagram} Iapetus is depicted as the circle moving on a prograde orbit, while the dust moves on retrograde orbits.  The white lines separate Iapetus' leading and trailing sides.  When orbits are circular (left), dust will solely darken the leading side, while the radial velocities of eccentric orbits allow dark material to reach part of the trailing side.}
\end{figure}

When particles have eccentric orbits, however, particle velocities are no longer perfectly azimuthal, and the radial components allow particles to strike the moon further along the equator (see Fig. \ref{extensiondiagram}).  Eccentricities induced by radiation pressure therefore provide a natural mechanism for extending dust coverage onto the trailing side.  

However, just as eccentricities act to extend coverage longitudinally, dust-orbit inclinations and Iapetus' varying orbital tilt should extend coverage latitudinally over the poles \citep[][and see Fig. \ref{Laplace}]{Burns96}.  Images of Iapetus, however, reveal bright, icy poles.  

As previously mentioned, thermal ice migration provides a mechanism for brightening the poles \citep{Spencer10}.  Icy patches on Iapetus darkened by exogenous dust increase in temperature as a result of their lowered albedo.  Sublimation rates, which depend exponentially on temperature \citep[e.g.,][]{Vyazovkin97}, thereby increase sharply.  This liberates bright ice and leaves behind an even darker surface.  The further darkened surface's temperature rises further, and the cycle repeats in a self-accelerating process until a lag deposit forms with thickness of order the thermal skin depth \cite{Spencer10}.  The result is that warm, darkened areas become extremely dark and ice-free, while the sublimed ice settles on the coldest areas of the moon---the trailing side and the poles.  

The distribution of dark material on the surface therefore holds several insights into ongoing processes on Iapetus as well as to the past and present prevalence of dark dust in the outer Saturnian system.  Unfortunately, it is difficult to observationally determine the dark layer's depth.  Bright-floored craters from small impactors that punctured through the dark layer constrain the layer to being much thinner than the crater's depth $\sim 10$ m \citep{Denk10}, while radar measurements \citep{Ostro06} imply Cassini Regio is on the order of decimeters deep.  

Given the above background, we wish to calculate the probability distribution for where on Iapetus dust would strike for three reasons:

a)  One can convert a probability distribution to a depth distribution (Sec. 3.2) and compare the resulting global map to the observed Iapetus surface.  Such a comparison tests the hypothesis that Iapetus is darkened by dust from Phoebe and can provide depths in areas where observations are not available.

b)  Calculated polar deposition rates of dust yield an estimate of the minimum sublimation rate required to overwhelm dust deposition and keep the poles bright.

c)  A global depth distribution provides the total volume of dark material on Iapetus.  This volume, coupled with the collision probabilities of dust calculated in Sec. 2, provides a probe of the total amount of dust collisionally generated in the outer Saturnian system over its history \citep[cf.][]{Bottke10}.

We subdivide this problem by first calculating the collision probability distribution over the surface of Iapetus in Sec. 3.1.  Then 3.2 converts this probability distribution to a depth distribution, and 3.3 estimates the sublimation rates required to keep the poles bright.  We postpone discussion of point c) to Sec. 5.

\subsection{Collision Probabilities as a Function of Latitude and Longitude}

\begin{figure}[!ht]
\includegraphics[width=9cm]{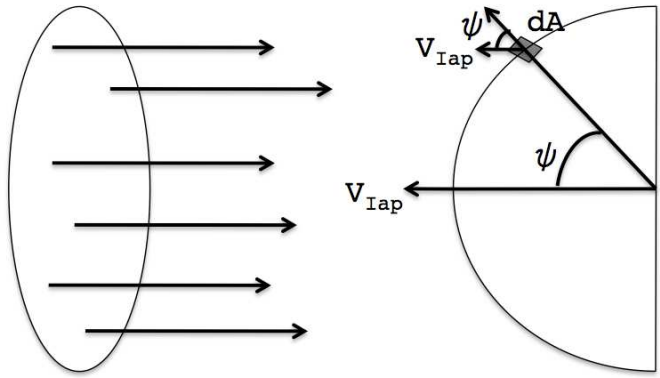}
\caption{\label{cos}  The orbits capable of striking Iapetus are well approximated by a uniform disk of parallel trajectories, shown on left.  Probabilities are then simply proportional to the projected area, given by $dA\: cos \psi$.  The apex of motion is at the leftmost point on the semicircle.}
\end{figure}

We now find the probability per unit area for particles striking Iapetus at latitude $\theta$ and longitude $\phi$.  Note that we can quickly determine the rough shape such a distribution should take in the limit of circular, uninclined orbits (a good approximation for large particles).  In this limit, dust particles strike Iapetus' leading side head-on.  Also, since $a_I$$\gg$$R_I$, we can approximate the orbits of Iapetus-striking particles as parallel straight lines.  Finally, in this approximation, our assumed uniform distribution in the variables $\Omega$ and $\omega$ for both orbits (see Sec. 2.2) translates into a uniformly distributed bundle of quasi-parallel trajectories capable of striking Iapetus.  In such a uniform field, the probability of an impact in a given area element simply is proportional to its projected area, given by $dA cos \psi$, where $\psi$ is the angle between Iapetus' velocity vector and the outward normal vector to the area element.  Equivalently, $\psi$ is the angular distance from the apex of motion (see Fig. \ref{cos}, in which Iapetus is moving to the left).  In this simple case then, the probability per unit area is a simple function of $\psi$,

\begin{equation}\label{oomp}
P(\theta, \phi) \propto cos\:\psi.
\end{equation}

This approximation is good over most of the leading hemisphere, though it is clearly incapable of describing the extension of the dark material onto the trailing side and of quantifying probabilities in the interesting transition region from the dark to the light terrains.  As described in Sec.~2.4, wrapping onto the trailing hemisphere cannot be the result of Iapetus' negligible gravitational focusing of retrograde particles (the maximum deflection of a retrograde particle by Iapetus' gravity is $\sim 1^{\circ}$).  Such extension is, however, a natural consequence of dust particles on eccentric orbits (see Fig. \ref{extensiondiagram}).  We therefore now calculate the probability distribution in the more general case of eccentric, inclined particles.

Following Greenberg's formalism (1982), we express the probability distribution function (pdf) as an integral over all the uniformly distributed angles ({\it $\Omega_I, \omega_I, \Omega_p, \omega_p$}), where the 'I' subscripts refer to Iapetus, and the 'p' subscripts to the dust particle.  Figure \ref{orbit} shows the geometry of an arbitrary orbit's three angular orbital elements $i, \Omega$ and $\omega$.  

As Greenberg (1982) notes, however, the problem's geometry does not depend on the values of $\Omega_I$ and $\Omega_p$ independently---the only geometrically meaningful quantity is their difference $\Delta \Omega$.  Furthermore, an important simplification can be made by approximating Iapetus' orbit as circular (its actual eccentricity = 0.03 and for circular dust orbits would lead to extensions of only $\sim 1^{\circ}$ onto the trailing side).  This obviates the need to specify $\omega_I$, the position of pericenter in Iapetus' orbit.  These two considerations reduce the phase space dimensionality from ({\it $\Omega_I, \omega_I, \omega_p, \Omega_p$}) to ({\it $\omega_p, \Delta \Omega$}), so that 
\begin{equation}\label{genrho}
\rho (\theta, \phi) = \int \rho (\theta, \phi, \omega_p, \Delta \Omega) \: d\omega_p \: d\Delta \Omega,
\end{equation}
where the integral spans the region in ({\it $\omega_p, \Delta \Omega$}) space in which collisions occur.  $\rho (\theta, \phi, \omega_p, \Delta \Omega)$ can further be expressed in terms of the conditional pdf $\rho (\theta, \phi)$ {\it given} the set ({\it $\omega_p, \Delta \Omega$}) multiplied by the probability density for ({\it $\omega_p, \Delta \Omega$}),
\begin{equation}\label{conditionalrho}
\rho (\theta, \phi, \omega_p, \Delta \Omega) = \rho (\theta, \phi | \omega_p, \Delta \Omega) \rho(\omega_p, \Delta \Omega),
\end{equation}
so
\begin{equation}\label{integralconditionalrho}
\rho (\theta, \phi) = \int  \rho (\theta, \phi | \omega_p, \Delta \Omega) \rho(\omega_p, \Delta \Omega) \: d\omega_p \: d\Delta \Omega.
\end{equation}
This simplifies the problem because $ \rho(\omega_p, \Delta \Omega)$, the probability of striking Iapetus (anywhere) given $\omega_p$ and $\Delta \Omega$, is already available \citep{Greenberg82}.  The problem is then reduced to finding $\rho (\theta, \phi | \omega_p, \Delta \Omega)$.

While analytically correct, Eq. \eqref{integralconditionalrho} is a formidable integral to compute numerically due to the scale separation in the problem.  The orbit is so large compared to the satellite that a minute change in $\omega_p$ or $\Delta \Omega$ shifts the location of impact drastically.  This sensitivity of $ \rho (\theta, \phi | \omega_p, \Delta \Omega)$ dictates extremely fine stepsizes in the integration.  When combined with the fact that the calculation must be done for $10^4$ separate timesteps, 8 initial conditions and 6 particle sizes, the scale separation indicates a brute force approach will be cumbersome at best.

\begin{figure}[!ht]
\includegraphics[width=9cm]{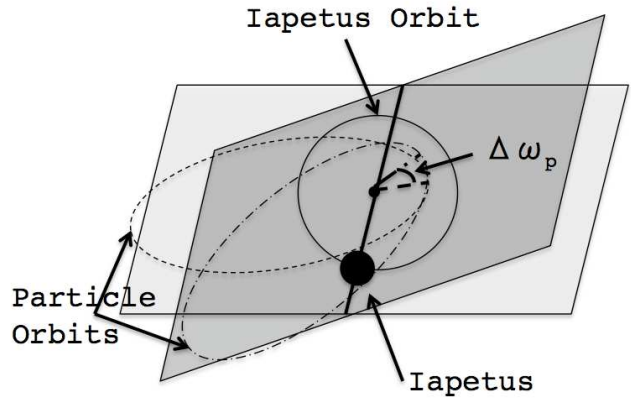}
\caption{\label{wp} Orbits can only cross along the line that marks the intersection of both orbital planes (the line of nodes).  Iapetus is depicted at one of the nodes, with its size greatly exaggerated.  For the particle orbits, there is a range in the angle from the node to pericenter ($\omega_p$) $\Delta\omega_p$ where collisions with Iapetus are possible.  Similarly there is a collisional range in $\Delta\Omega$, the angle that rotates the line of nodes in the plane (not shown).}
\end{figure}

However, this approach considers each orbital orientation individually.  The scale separation lets us consider well-defined {\it groups} of orientations.  Consider an orientation where the orbits cross exactly, i.e., the particle would pass through the center of Iapetus.  There is a range in $\Delta \Omega$ and $\omega_p$ around this orbital orientation where the orbits no longer exactly cross but are still close enough that the particle impacts Iapetus (Fig. \ref{wp}).  

As previously argued, the facts that $a_I$$\gg$$R_I$ and that gravitational focusing is negligible mean that, near impact, we can approximate these orbits as a uniformly distributed disk of parallel trajectories.  This approach allows one to coarsen stepsizes while retaining the symmetries in the problem and maintaining the fidelity of the final distribution.

With these considerations, the problem of calculating the distribution function is more tractable.  Our approach will be to first find the latitude and longitude for exactly crossing orbits, and then to find how the disk of parallel orbits around the central orbit maps onto the spherical surface of Iapetus.  

Given a particular $\Delta \Omega$, one can combine it with the inclinations to determine the orientation of the orbital planes relative to each other \citep[see Fig. 2 in][]{Greenberg82}.  The relative inclination $i'$ is given by spherical trigonometry,
\begin{equation}\label{cosiprime}
cos\:i' = cos\:i_I\:cos\:i_p + sin\:i_I\:sin\:i_p\:cos\: \Delta \Omega.
\end{equation}
Within the particle's orbital plane, $\omega_p$ sets the orientation of the orbit.  Having approximated Iapetus' orbit as circular, we do not have to consider the satellite's orientation within its orbital plane.  From Fig. \ref{wp}, it is clear that most particle-orbit orientations do not result in a crossing.  Furthermore, if the two orbits are to cross, they must do so at either of the two nodes where Iapetus' orbit pierces the mutual line of nodes.  

The facts that the orbits must cross at a node, and that those respective points on the orbits must therefore be equidistant from Saturn sets the possible values of ${\it \omega_p}$ (note that this would not be the case if both orbits were substantially non-circular).  The angle from pericenter to the ascending node is, by definition, ${\it -\omega_p}$.  From the equation for an ellipse, we therefore have the condition,
\begin{equation}\label{aI}
a_I  = \frac{a_p(1 - e_p^2)}{1 + e_p\:cos(-\omega_p)}.
\end{equation}
Rearranging,
\begin{equation}\label{coswp}
cos\:\omega_p = \frac{1}{e_p} \Bigg[\frac{a_p(1 - e_p^2)}{a_I} - 1\Bigg].
\end{equation}
Since cosine is an even function, Eq. \eqref{coswp} gives two solutions $\pm {\it \omega_p}$, reflecting the ellipse's symmetry across its long axis.  Similarly, another pair of solutions are present at the descending node, located at an angle of 180 - $\omega_p$ from pericenter.  In general, therefore, four orientations yield crossing orbits for a given $\Delta \Omega$ \citep[see Fig. 1 in ][]{Wetherill67}.

The imposed circularity of Iapetus' orbit means that these four orbits will strike symmetrically about the equator and about the longitude that corresponds to the apex of motion (this occurs because the satellite rotates synchronously).  In other words, if we define the longitude in the direction of motion as $0^{\circ}$, the four exactly-crossing orbits will strike at ($\pm \theta, \pm \phi$).  These can later be adjusted to conform with the conventional longitude in the direction of motion---the zero-longitude meridians are currently under revision for Saturn's satellites \citep{Roatsch09}.  Given a $\Delta \Omega$, one therefore only has to compute ($\theta, \phi$) for one of the four orientations and then straightforwardly substitute for the other three.  Here we choose to consider collisions at the ascending node.

\begin{figure}[!ht]
\includegraphics[width=9cm]{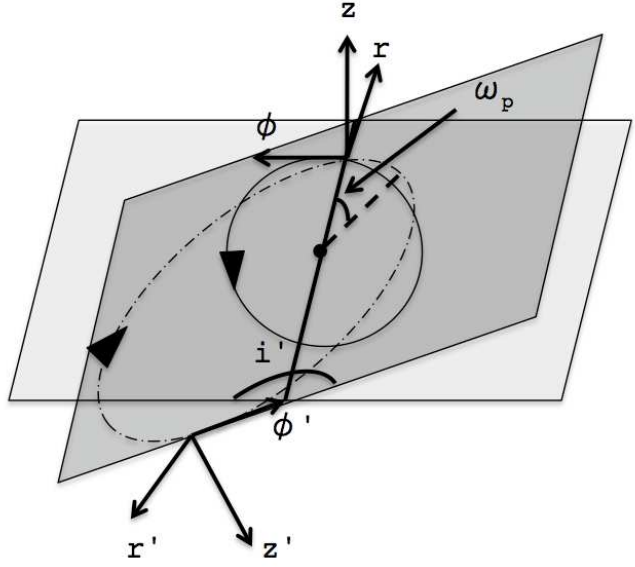}
\caption{\label{system}  Iapetus' circular orbit is executed in the lighter horizontal plane, while the particle's orbit is carried out in the inclined darker plane, with the two crossing at the particle orbit's ascending node.  We choose to work in a cylindrical coordinate system centered at the ascending node where the z direction is Iapetus' orbit normal.  For simplicity, we first express the particle's velocity in its own orbital plane, where z' is the orbit normal.  The relative inclination $i'$ and argument of pericenter $\omega_p$ are also depicted.}
\end{figure}

The relevant vector to consider is the relative velocity vector in a coordinate system centered on the ascending node (see Fig. \ref{system}).  

The spherical angles that define the direction of the particle's relative velocity vector in this system determine the location of impact on the Iapetus surface.  The relative velocity vector is given by
\begin{equation}\label{relvel}
\vec{\bf{v}}_{rel} = \vec{\bf{{v}}}_p -  \vec{\bf{{v}}}_I.
\end{equation}
Having approximated Iapetus' orbit as circular, $\vec{\bf{{v}}}_I$ is always azimuthal.  $\vec{\bf{{v}}}_I$ can therefore be simply written in terms of the uniform circular velocity,
\begin{equation}\label{vI}
\vec{\bf{{v}}}_I = \sqrt{\frac{GM_{Sat}}{a_I}} \Bigg(\begin{array}{c}0\\1\\0\end{array}\Bigg).
\end{equation}
$\vec{\bf{{v}}}_p$ is given in the particle's orbital plane in cylindrical coordinates by \cite{Hamilton93},
\begin{equation}\label{vporbitalplane}
\vec{{\bf v}}_p' =  \sqrt{\frac{GM_{Sat}}{a_p(1-e_p^2)}}\Bigg(\begin{array}{c}e_psinf \\ 1 + e_pcosf \\ 0 \end{array}\Bigg).
\end{equation}
Since we are interested in the particle's velocity at the ascending node in particular, we plug in ${\it f = -\omega_p}$ (see Fig. \ref{system}).  At the ascending node, the unit vectors $r$ and $r'$ align, but the remaining unit vectors are misaligned by the relative inclination $i'$.  We therefore rotate $\vec{\bf{{v}}}_p$ by an angle $-i'$ around the $r$ axis (see Fig. \ref{system}), yielding
\begin{equation}\label{vpascendingnode}
\vec{{\bf v}}_p= \sqrt{\frac{GM_{Sat}}{a_p(1-e_p^2)}}\Bigg(\begin{array}{c}-e_psin\omega_p \\ cosi' \:(1 + e_pcos\omega_p) \\ sini'(1 + e_pcos\omega_p)\end{array}\Bigg).
\end{equation}
The relative velocity vector in Eq. \eqref{relvel} can then be obtained from Eqs. \eqref{vI} and \eqref{vpascendingnode}, plugging in for $cos \omega_p$ from Eq. \eqref{coswp}.  The latitude $\theta$ and longitude $\phi$ on Iapetus where the particle strikes are then given by,
\begin{eqnarray}
\theta = Lat = -tan^{-1}\Bigg(\frac{v_z^{rel}}{\sqrt{(v_x^{rel})^2 + (v_y^{rel})^2}}\Bigg), \\
\phi = Long = tan^{-1}\Big(\frac{v_y^{rel}}{v_x^{rel}}\Big) - 180^{\circ}.
\end{eqnarray}
The symmetry described earlier can then be used to find the latitudes and longitudes for the other three orientations.  We have thus determined the location where particles on the four possible crossing orbits (for a given $\Delta \Omega$) would impact.  The last piece is to include the disk of parallel trajectories around these crossing orbits that can still impact Iapetus (see Fig. \ref{wp}).

As mentioned earlier, the scale separation in the problem allows us to consider all the orbits that can strike Iapetus close to the crossing orbit as parallel lines with a uniform probability distribution.  The situation is analogous to the one presented at the beginning of the section, except with the $\psi = 0$ direction now interpreted as the incoming trajectory at latitude and longitude $\theta$ and $\phi$, respectively.  The probability is again proportional to the projected area, so normalizing the probability distribution we obtain
\begin{equation}\label{normP}
P(\psi) = \frac{cos\psi}{\pi {R_I}^2},
\end{equation}
which falls to 0 as $\psi$ reaches $90^{\circ}$ like it should.  Obviously this only applies to the hemisphere facing the disk---for wherever $\psi > 90^{\circ}$, P($\psi$) = 0.  Again, this would not be the case with substantial gravitational focusing, but as a result of the high relative velocities due to dust orbits being retrograde, gravitational focusing is negligible (Sec. 2.4).  The probability $P(\psi)$ can then be straightforwardly converted to a probability per $d\theta$ and $d\phi$ for substitution into Eq. \eqref{integralconditionalrho}.

This provides a prescription for numerically computing $\rho(\theta, \phi)$, the probability density function that we originally set out to find, as a function of latitude and longitude .  Cycling over $\Delta \Omega$, at each step, we identify the four crossing orbits and their associated probabilities within the interval, using \cite{Greenberg82}.  Then, for each of the four crossing orbits, we "spread" the respective probability across the hemisphere defined by the crossing orbit through the distribution in Eq. \eqref{normP}.  

Note that, since $\rho(\theta, \phi)$ depends implicitly on particle eccentricities (cf. Eq. \ref{vpascendingnode}), it will also be a function of particle size.  We can write this explicitly, and straightforwardly convert $\rho(\theta, \phi)$ to a probability per unit area, by defining $P(r, \theta, \phi) = \rho(r, \theta, \phi) / cos(\theta){R_I}^2$.  We choose to use $P(r, \theta, \phi)$, the normalized probability per unit area, in subsequent calculations.

\subsection{Calculating Depths}

\begin{figure}[!ht]
\includegraphics[width=9cm]{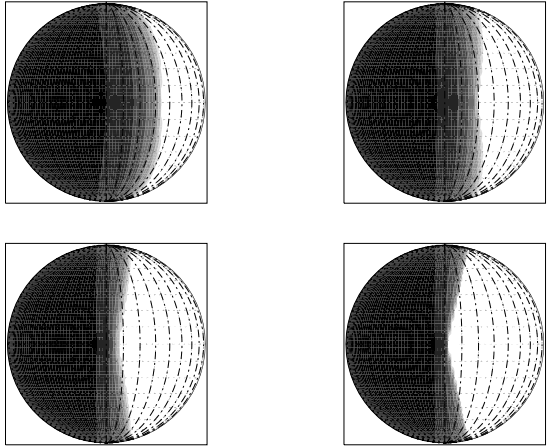}
\caption{\label{sizeDistributions} Moving from the top-left figure clockwise, probability density functions for 5, 10,  500 and 50 $\mu$m particles.  Plots represent equatorial views where the vertical line in the center represents the boundary between leading and trailing sides.  As such, the apex of motion is at the leftmost point on each figure.  Contours represent successive 10-fold decays from the peak value at the apex of motion, down to $10^{-7}$ of the apex value.  Dot-dashed lines are drawn every $10^{\circ}$ in longitude.}
\end{figure}

Since radiation pressure produces different orbital histories and is particle-size dependent, different particle sizes have different pdfs.  Fig. \ref{sizeDistributions} shows the probability density functions for 5, 10, 50 and 500 $\mu$m particles, with each contour representing a successive 10-fold decay from the peak value at the apex of motion at the left-most point of each figure.

The figure shows that smaller particles extend farther onto the trailing side near the equator.  This is due to their higher eccentricities, as discussed at the beginning of Sec. 3 (see Fig. \ref{extensiondiagram}).  The distributions for particles $\gtrsim 50 \mu$m quickly converge to the large-particle limiting distribution, depicted for $500 \mu$m particles.  The eccentricities of these larger particles are too low to cause them to significantly wrap around the equator onto the trailing side; however, the coverage over the poles is dominated by the precession of Iapetus' orbit, which is independent of particle size.  

We now use such probability density distributions $P(r, \theta, \phi)$ to estimate the depth of dust as a function of position on Iapetus' surface.  The volume of dust particles within $dr$ of size $r$ that lands within an area $A$ on Iapetus at latitude $\theta$ and longitude $\phi$ can be expressed as
\begin{equation}\label{volume}
Volume(r, \theta, \phi) = N(r)\times P(r, \theta, \phi) \times A \times V(r),
\end{equation}
where $N(r)$ is the number of dust particles within $dr$ of radius $r$ generated in the outer Saturnian system and $V(r)$ is the volume of a spherical particle of radius $r$.  Unfortunately, the current (or past) particle size distribution N(r) is not well constrained observationally.  We therefore consider a variety of exponents for distributions of the form, 
\begin{equation}\label{powerlaw}
N(r) = Dr^{-\beta} dr,
\end{equation}
where $D$ is a normalization constant, and $\beta$ is the (negative) power law index of the particle size-frequency distribution.  Finally, the depth can be estimated (to within a packing efficiency factor) as the volume over an area element divided by the area of the surface element.  

The final integration over the range of particle sizes to find the total dust depth is complicated by the fact that larger particles, being less affected by Poynting-Robertson drag, take longer to reach Iapetus from Phoebe.  
We can consider two limiting cases:  

a)  Most of the debris in the outer Saturnian system was generated early in the Solar System's history (i.e., the mass contribution from the Phoebe ring is negligible).  In this limiting case, all the particles with Iapetus-collision timescales (and destruction lifetimes) smaller than the age of the Solar System will have had time to impact Iapetus and collision timescales across different particle sizes are irrelevant.

b)  The mass in the outer Saturnian system has been generated at a constant rate over its history.  In this case where particles are continuously resupplied, smaller particles that decay inward faster will have a larger effect than they would have in the first case.

We begin by considering case a) where we investigate all particles with collision timescales $\tau_C$ smaller than the age of the Solar System on an "even footing."  Since dust particles' semimajor axes decay exponentially through Poynting-Robertson drag on a timescale $\tau_{PR}$, and since the ratio of Phoebe's to Iapetus' semimajor axes is $\approx 3.6$, particles that strike Iapetus do so on roughly a single e-folding timescale, i.e., $\tau_{C} \sim \tau_{PR}$.  The P-R timescale is given by Eq. \eqref{pr}.  This implies that the largest dust size to consider is $\sim 1$ cm, with corresponding $\tau_{PR} \sim 1$ Gyr.

Integrating over all particle sizes,
\begin{equation}\label{depth}
Depth(\theta, \phi) \propto \int_{r_{min}}^{r_{max}} r^{3 - \beta} \times P(r, \theta, \phi) dr.
\end{equation}
For $r_{min}$, we use the smallest size of long-lived particles from Phoebe, approximately 5 $\mu$m (see Sec. 2.1).  At the other limit, we use $r_{max}$ $\sim 1$ cm.  Should lifetimes from catastrophic collisions between particles or other processes \citep[see][]{Burns01} be lower than $\sim 1$ Gyr, $r_{max}$ must be considered more carefully.  

We now address case b), where particles are continuously resupplied.  In this circumstance we can consider each particle size to fall onto Iapetus at a characteristic rate,
\begin{equation}\label{rate}
Rate(r, \theta, \phi) = \frac{Volume(r, \theta, \phi)}{\tau_{C}},
\end{equation}
where $\tau_{C}$ is the characteristic collision timescale and is $\sim \tau_{PR}$.  We can express the depth then as
\begin{equation}\label{depthrate}
Depth(r, \theta, \phi) \propto \frac{Rate(r, \theta, \phi) \times t}{A},
\end{equation}
where $t$ is the interval over which dust has been accumulating.  From Eq. \eqref{pr}, we find that $\tau_C \propto r$, so plugging in for the volume as was done in the first case, we find that 
\begin{equation}\label{depthgamma1}
Depth(\theta, \phi) \propto \int_{r_{min}}^{r_{max}} r^{2 - \beta} \times P(r, \theta, \phi) dr.
\end{equation}

A comparison between Eqs. \eqref{depth} and \eqref{depthgamma1} shows that a constant rate of dust production simply acts to steepen the effective power-law index, because small particles will arrive at Iapetus more quickly than large ones.  The effective power-law index will be intermediate between the limiting cases of Eqs. \eqref{depth} and \eqref{depthgamma1}, and the depth can therefore be generally expressed as

\begin{equation}\label{depthGen}
Depth(\theta, \phi) \propto \int_{r_{min}}^{r_{max}} r^{3 - (\beta + \gamma)} \times P(r, \theta, \phi) dr,
\end{equation}
where $\gamma$ is a number between 0 and 1 that parametrizes the constancy of dust production over the age of the Solar System.  A $\gamma$ of 0 and 1 would therefore, respectively, correspond to cases (a) and (b) introduced at the beginning of this section (3.2).

The constant of proportionality in Eq. \eqref{depthGen} is {\it a priori} highly uncertain.  An important, though poorly constrained, quantity is the time by which Iapetus had become tidally locked to Saturn.  Iapetus' dichotomy could not have formed prior to this time, as a non-synchronously rotating Iapetus would receive dust equally on all sides.  Furthermore, the thermal models required to explain its sharp albedo boundaries and bright poles require Iapetus' slow 79 day synchronous period \cite{Spencer10}.  \cite{Castillo07} estimate tidal locking occurred between 200 Myr and 1 Gyr after formation.  Moreover, \cite{Bottke10} argue that most of the dust in the outer Saturn system should have been generated in the first few 100 Myr.  Case a) reflects a situation where most of the dust in the outer Saturn system was generated early and Iapetus was able to quickly achieve synchronous rotation so that this dust mass arrived after locking.  If, on the other hand, the timescale for tidal evolution is long ($\sim$ 1 Gyr), one might expect the production of the relevant dust to be fairly constant---case b)---since any initial flurry of dust (should there have been one) would have arrived too soon.  Despite these uncertainties, one can still normalize the depths over the surface {\it a posteriori} through a measurement of depth at a particular position $(\theta, \phi)$.  

In studying Iapetus with the Cassini radar instrument, \cite{Black04} found little hemispheric asymmetry in albedo at a wavelength of 13 cm, while \cite{Ostro06} observed a strong dichotomy at 2 cm wavelength.  \cite{Ostro06} interpret these results as implying contamination of ice with dark material to a depth of one to several decimeters.  The latter's measurement on the leading side of Iapetus was centered on ($66^\circ$ W, $+39^\circ$ N), but the beam size was comparable to the angular size of the satellite (beam/R = 1.36).  

\begin{figure}[!ht]
\includegraphics[width=9cm]{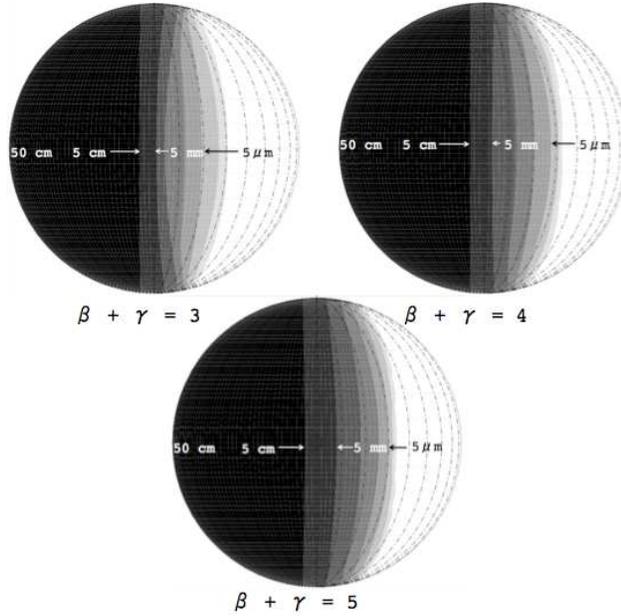}
\caption{\label{bplusg} Depth contours representing 10-fold decays from the peak value (at the extreme left of the figure) for $\beta_{eff} = \beta + \gamma = 3, 4$ and $5$, {\it assuming} a peak depth at the apex of motion (extreme left of each figure) of 0.5 m.  Plots represent equatorial views where the vertical line in the center represents the boundary between leading and trailing sides.  Dash-dotted lines are drawn every $10^{\circ}$ in longitude.}
\end{figure}

\begin{figure}[!ht]
\includegraphics[width=9cm]{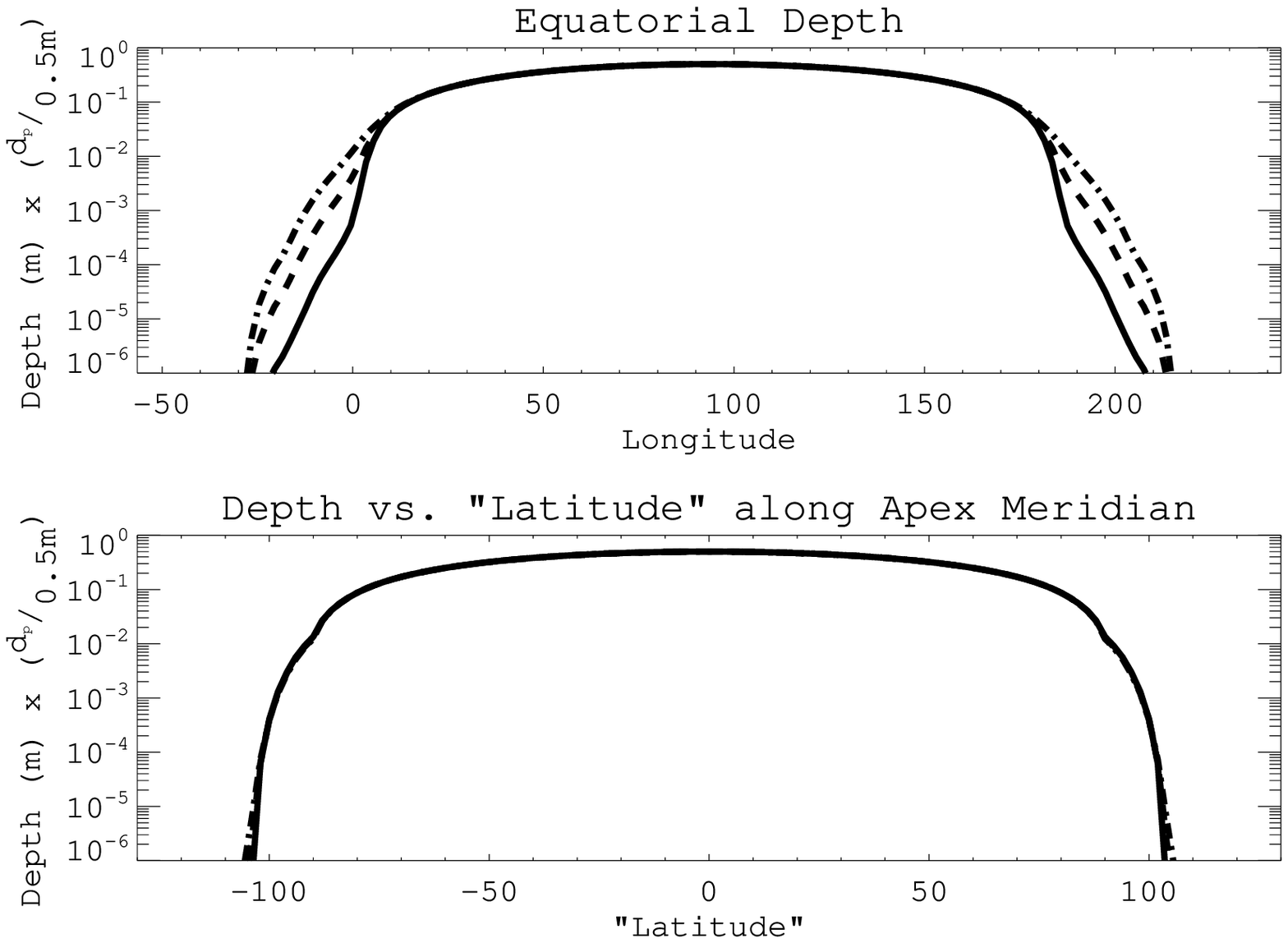}
\caption{\label{eqMerCuts} Top graph shows depth vs longitude along the equator for $\beta_{eff} = \beta + \gamma = 3$ (solid), $\beta_{eff} = 4$ (dashed), $\beta_{eff} = 5$ (dash-dotted).  Bottom graph shows depth vs latitude along the meridian passing through the apex of motion (longitude $\sim 90^{\circ}$ W).  The concept of latitude has been extended beyond $\pm 90^{\circ}$ along the corresponding meridian on the trailing side of Iapetus to show the extension of dark material over the poles.}
\end{figure}

While this measurement is imprecise, it does set the order of magnitude of the dark material's depth.  Figure \ref{bplusg} shows depth contours for three different choices of effective power law index, $\beta_{eff} \equiv \beta + \gamma$ in (25), {\it assuming} a peak depth of dust at the apex of motion (extreme left of each figure) of 0.5 m.  Figure \ref{eqMerCuts} provides depths following the equator and meridian passing through the apex of motion at $\sim 90^{\circ}$ W for the same cases.

The top graph in Fig. \ref{eqMerCuts} shows that one should expect extension of dark material $\sim 20-30^{\circ}$ onto the trailing side for all expected particle-size distributions.  Only small particles ($\lesssim 25 \mu$m in size), having more eccentric orbits, can significantly reach onto the trailing side.  As a result, the shallowest effective power-law index ($\beta_{eff} = 3$), having fewer small particles, yields a spatial distribution that extends onto the trailing side significantly less.  

The bottom graph shows the extension over the poles.  Far from Saturn, solar torques dominate torques from Saturn's oblateness and cause orbits' angular momentum vectors to precess, keeping the inclination roughly constant.  Because the inclination is set by initial conditions (i.e., Phoebe's orbital inclination), and is independent of particle size, the graphs for all three power-law indices overlap.  The extension over the poles (``latitudes" $> \pm 90^{\circ}$ along the meridian onto the trailing side) is due to both particle-orbit inclinations and Iapetus' orbital precession.  

One should be careful in distinguishing measured depths of dark material (through radar or otherwise) from depths of dust accumulated over Iapetus' history.  If the model of dust deposition and subsequent thermal ice migration is correct, the depth of dark material would be the sum of the contributions from exogenous dust and from the native lag deposit (see discussion in Sec. 3).  As long as the depth measured is substantially larger than the expected depth of the lag deposit, the distinction is minor.  Figure \ref{bplusg} assumes a peak depth {\it of dust} of 50 cm--with no impact-gardening, this would imply an actual depth of dark material about $20\%$ greater (with a 10 cm lag deposit).  Should improved measurements of the peak dust depth become available, the contours on these maps could be straightforwardly rescaled.  

The additional lag-deposit depths are not included in our modelling; however, since exogenous dust acts as the trigger to thermal ice migration, the maps above should be good tracers (at low latitudes) of which areas will be dark and which will be light.  We can therefore attempt to predict the boundary of Cassini Regio.  

While the figures show a rapid fall-off in depth on the trailing side, maps of Iapetus show no gradation in albedo.  Areas initially darkened by infalling dust and receiving strong insolation become almost completely blackened.  As a result, the equatorial regions of the leading side appear uniformly dark, while dust-free areas of the trailing side and the colder poles, where the ice that sublimated at lower latitudes settles, appear about ten times brighter.  

\begin{figure}[!ht]
\includegraphics[width=9cm]{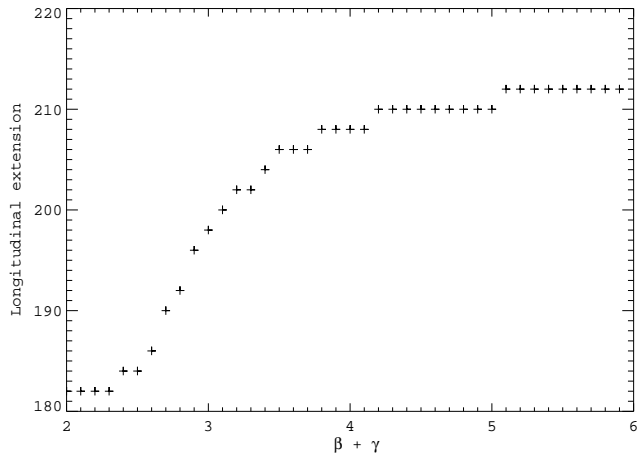}
\caption{\label{extension} The extension in longitude onto the trailing side, chosen as the longitude at which the depth falls below $10^{-5}$ the peak value, for different values of $\beta_{eff} = \beta + \gamma$.  The discrete steps are the result of the resolution of the calculation---$2^{\circ}$ in longitude. }
\end{figure}

Thus, if one ignores the contours, the plots above argue for a blackened leading side extending between twenty to thirty degrees in longitude onto the trailing side for $3 < \beta_{eff} < 5$.  In fact, this holds true for all $\beta_{eff} > 3$ as shown in Fig. \ref{extension}.  At low latitudes, this matches maps of Iapetus' albedo well(see Fig.\ref{iapetusmap}).  Deposition of exogenous dust therefore neatly explains the boundaries of the dark material at low latitudes for the entire range of likely power-law indices (accordingly, Spencer $\&$ Denk's (2010) model explains the sharp boundaries in albedo as well as the bright poles).  

If indeed Iapetus was initially darkened by dust from the outer Saturnian system, the extension onto the trailing side seems to exclude the shallowest power-law indices in the particle size distribution, $\beta+ \gamma < 3$.  While it is encouraging that all $\beta_{eff} > 3$ are consistent with the observed distribution, the small slope in Fig. \ref{extension} for $\beta_{eff} > 3$ renders the longitudinal coverage on Iapetus a comparatively poor indicator of the responsible particle size distribution.  

\begin{figure}[!ht]
\includegraphics[width=9cm]{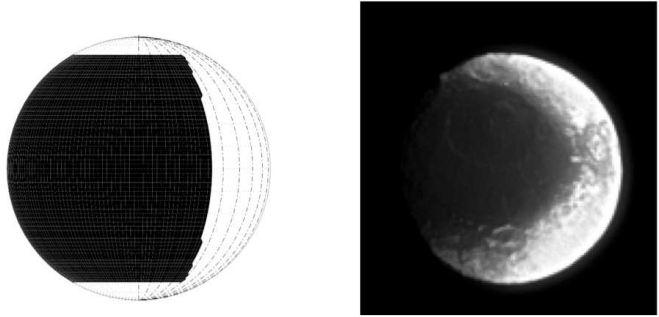}
\caption{\label{iapetusfinal} Model of the Iapetus surface assuming a peak depth of 50 cm, $\beta = 3.5$ and $\gamma = 1$.   All areas with depths $> 5 \mu$m have been uniformly darkened and the poles beyond $\pm 60^{\circ}$ latitude brightened to artificially account for thermal ice migration.  On the right is an image taken by Cassini at roughly the same orientation for visual comparison (obtained from the Planetary Photojournal---PIA08273).}
\end{figure}

Fig. \ref{iapetusfinal} shows the distribution for $\beta = 3.5$, the power-law index for an idealized infinite collisional cascade \citep{Burns01}.  It assumes a constant supply of particles ($\gamma = 1$) and artificially accounts for thermal ice migration by brightening the poles down to the observed latitude of $\sim \pm 60^{\circ}$ and by completely darkening areas with depths greater than $5 \mu$m.  

As previously mentioned, $5 \mu$m particles are the smallest Phoebe-generated particles that would strike Iapetus; therefore, this boundary for the anticipated depth is roughly where one should expect to transition from uniform darkness to the stochastic dalmatian patterns observed in the closest Cassini flyby of Iapetus \citep{Denk10}.  This seems consistent with a visual inspection of maps (Fig. \ref{iapetusmap} and see also \citep{Blackburn10}) and images of Iapetus by Cassini like the one shown alongside in Fig. \ref{iapetusfinal}.

An important difference between the modeled surface and the observed distribution is that the theoretically derived dark terrain is concave, in the sense that if one follows a meridian on the boundary, one sees that the dark material extends further in longitude at higher latitudes.  Cassini Regio is convex, as can be clearly seen in Fig. \ref{iapetusmap}, which uses a simple cylindrical projection where meridians appear as vertical lines.  Since temperatures should drop smoothly as one moves from equator to pole, perhaps the discrepancy results from thermal ice migration.  Indeed, such models \citep{Spencer10} are able to reproduce this concavity.

Apart from the long-known albedo dichotomy on Iapetus, \cite{Denk10} recently detected a new {\it color} dichotomy on Iapetus in which the leading side of Iapetus is substantially redder than the trailing side.  The color dichotomy seems to extend farther pole-ward, and transitions more gradually onto the trailing side.  Perhaps, as \cite{Denk10} point out, the color dichotomy more faithfully traces where the dust landed while the albedo dichotomy reflects thermal ice migration's modification of the initial pattern of dust deposition.  Further work is needed to ascertain quantitative agreement between observations of the color dichotomy and theoretical models like those presented here.

\subsection{How Much Sublimation Is Required to Paint the Poles Bright?}

Substantial amounts of dust should have struck Iapetus at high latitudes; however, the poles appear bright (Fig. \ref{iapetusmap}).  If the preceding section is correct, this means that bright, sublimed ice from lower latitudes must be settling on the polar regions faster than dark dust is landing on them.  Our collisional flux then provides an opportunity to constrain sublimation rates.

We found in the previous section that the distribution at the boundary between the leading and trailing sides depends on the underlying particle size distribution of dust; however, farther from the boundaries, as argued at the beginning of Sec. 3.1, the depth (at low latitudes) should scale approximately as $cos\psi$ {\it independent} of the particle size distribution, where $\psi$ is the angular distance from the apex (cf. Eq. \eqref{normP}).

Therefore, following the meridian that passes through the apex of motion (longitude $\approx 90^{\circ}$), at a latitude of $60^{\circ}$ the depth should be roughly half that at the apex ($cos 60^{\circ} = 1/2$).  This point in the polar region would receive more dust than any other point at the same latitude as it has the minimum angular distance from the apex.  The fact that this point in the polar region with maximum dust flux appears bright provides the strongest constraint on the minimum sublimation rate required to keep the poles bright.  Assuming a peak depth of dust at the apex of 50 cm as done in the previous section based on radar measurements \citep{Ostro06}, this implies an average rate for the minimum polar dust deposition of $\sim 25 cm / 5 Gyr$ or $\sim 50 \mu m/Myr$.

It is possible, however, that the average rate of deposition would not match the rate of sublimation.  Maybe in the past \citep[when deposition rates of dust in the outer Saturnian system were likely higher,][]{Bottke10}, the poles of Iapetus were dark.  Perhaps only recently did deposition of ice exceed that of dust and hide evidence of past dark poles.  In that case the rate of ice sublimation required to keep the poles bright would be lower than the average dust deposition rate.  Dark-ringed craters in the polar terrains could support such a conjecture, but current observations are unable to distinguish between these two general possibilities.

Therefore, given only the observation that Iapetus' poles are bright {\it today}, we now try to roughly constrain the current sublimation rate.  We can estimate the deposition rate of the material coming from the Phoebe ring using its measured optical depth. 

For lack of better information, we assume that the entire volume of the ring has the same particle size distribution.  In this case, the rate of dust deposition at a latitude of $60^{\circ}$ along the meridian passing through the apex (longitude $\sim 90^{\circ}$ W) can be directly obtained from Eqs. \eqref{rate}, \eqref{volume}, \eqref{powerlaw} and \eqref{pr},
\begin{equation}\label{poleRate}
Rate(60^{\circ}, 90^{\circ}) \sim \int_{r_{min}}^{r_{max}} \frac{Dr^{-\beta} \: V_d \: P(r, 60^{\circ}, 90^{\circ}) \: \frac{4}{3}\pi r^3}{0.1\: r \: \frac{Myr}{\mu m}} \: dr,
\end{equation}
where $V_d$ is the volume of the disk and $P(r, 60^{\circ}, 90^{\circ}$) is the probability per unit area for dust striking at the longitude corresponding to the apex ($\sim 90^{\circ} W$) and a latitude of $60^{\circ}$.  In pursuing an order of magnitude estimate, we take all particle sizes in the Phoebe ring (i.e., $\gtrsim 5 \mu$m) to have probability $\sim 1$ of striking Iapetus (cf. Fig. \ref{probabilities}).  Furthermore, we consider the depth distribution to be simply proportional to $cos \psi$ (Fig. \ref{cos}), with depths of 0 on the trailing side.  For the integrated probability to yield unity, $P(r, 60^{\circ}, 90^{\circ}) \approx 3\e{-7}$ km$^{-2}$.

We estimate $V_d$ as the volume of a disk 300 $R_S$ in radius and 40 $R_S$ thick \cite[as done by][]{Verbiscer09}, or $\sim 2 \times 10^{21}$ km$^3$.  This yields

\begin{equation}\label{numPoleRate}
Rate(60^{\circ}, 90^{\circ}) \sim 10^{16} \frac{km \: \mu m}{Myr} \int_{r_{min}}^{r_{max}} Dr^{2 - \beta} dr.
\end{equation}

The final integral here is provided by the definition of the normal optical depth,

\begin{equation}\label{optdepth}
\tau = \int_{l_{min}}^{l_{max}} \int_{r_{min}}^{r_{max}} n(r,l) Q_{ext} \sigma (r) dr dl,
\end{equation}
where $Q_{ext} = Q_{abs} + Q_{scat}$.  Following \cite{Verbiscer09}, we take values of $Q_{abs} = 0.8$ and $Q_{scat} = 0.2$.  They estimate $\tau \sim 2 \e{-8}$.  Making again the simplifying assumption that the number density does not depend on the distance along the line of sight $l$ (trivializing the integration over $l$),
\begin{equation}\label{numOptDepth}
\frac{2\e{-8}}{40 R_S \: \pi} = \int_{r_{min}}^{r_{max}} D r^{2 -\beta} dr.
\end{equation}

If we are interested in the rate of deposition due to the material currently seen in the Phoebe ring (i.e., if we take the same limits of integration), the integral can then be plugged into Eq. \eqref{numPoleRate}, yielding a rate of dust deposition at the leading edge of the polar region of $\sim 50 \mu m / Myr$, which is of the same order as the average deposition rate calculated earlier.  This estimate also agrees with that given by \cite{Verbiscer09} of $\sim 40 \mu$m / Myr.  

In either event, Iapetus must today be actively depositing on the order of tens of $\mu$m per Myr of sublimed ice onto the poles in order to keep them bright.  This ice could originate from either the leading or trailing side.  In Cassini Regio, with daytime temperatures of 130 K, ice will sublime over extremely short timescales---about 1000 $\mu$m in 8000 years \citep{Spencer10}.  Very quickly, a dark layer thicker than the thermal skin depth will form, making it difficult for further ice to sublimate.  In this case, the rate of impact gardening, which brings fresh ice to the surface, will determine the rate of sublimation from Cassini Regio.  Unfortunately, impact gardening depths on Iapetus are not well constrained \citep{Spencer10}.

The temperature on the brighter trailing side indicates a sublimation rate of about $100 \mu m / Myr$ \citep{Spencer10}.  Given that the area of the trailing side at latitudes $< 60^{\circ}$ is comparable to that of both polar regions combined, this sublimation rate seems capable of overwhelming polar dust deposition.  Further data on the Phoebe ring will help further constrain the required sublimation rates, and improved modeling of the surface processes on Iapetus will ultimately dictate the consistency of these two pictures.

\section{Dust from other Irregular satellites}

A long-standing objection to Soter's model of dust infall from Phoebe has been that Cassini Regio's spectrum differs from that of Phoebe \citep[e.g.,][]{Tholen83, Cruikshank83, Buratti02}.  One possible resolution is that the surface compositions are in fact similar, but other effects such as Rayleigh scattering \citep{Clark08}, cause the spectra to differ.  Another possibility is that Iapetus has also been coated by dust from the irregular satellites other than Phoebe \citep{Buratti05, Tosi10}.  \cite{Grav03} and \cite{Buratti05} find that many of these new irregular satellites have a reddish color similar enough to that of Hyperion and Cassini Regio to suggest a link between them.  

Unlike the regular satellites (ignoring Iapetus) that move on low-eccentricity orbits close to Saturn's equatorial plane, the irregulars have widely varying inclinations and eccentricities.  This implies a violent collisional history between irregular satellites as differing precession rates would have led to crossing orbits and consequent collisions \citep{Nesvorny03}.  By modeling this process of collisional grinding numerically, \cite{Bottke10} estimate that $\sim 10^{20}$ kg of dust should have been generated in the outer Saturn system, particularly early in the Solar System's lifetime.  

\begin{figure}[!ht]
\includegraphics[width=9cm]{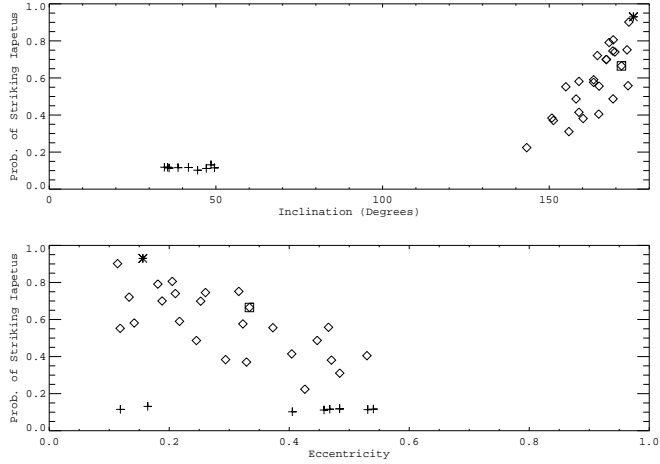}
\caption{\label{irr} Collision probability with Iapetus for $10 \mu$m grains that start with the orbits of today's irregular satellites.  Plus signs represent prograde irregulars, while open diamonds are retrograde.  Inclinations are measured relative to Saturn's orbital plane.  The asterisk represents Phoebe, and the boxed diamond Ymir (of importance below).}
\end{figure}

We therefore explore the likelihood that debris from these other irregular satellites would collide with Iapetus.  Fig. \ref{irr} shows the probability that $10 \mu$m grains will strike Iapetus if they start with the orbits of the various irregular satellites known today, plotted against both today's value of the parent-satellite orbit's inclination and eccentricity.  The orbits of the current irregular satellites are not chosen to be necessarily representative of their orbits over the course of the Solar System's history---the irregular satellites seen today are likely the fragments of past satellites and are subject to increasingly strong gravitational perturbations from the Sun the further out in Saturn's Hill sphere they reside \citep{Nesvorny03, Turrini08, Bottke10}.  Rather, we chose the current orbits as a way of sampling the orbital phase space of irregular satellites.  

The efficiency with which material is supplied to Iapetus differs markedly between the prograde (plus signs) and retrograde (diamond) satellites (Fig. \ref{irr}).  This can be understood through a simple particle-in-a-box estimate of the collision timescale, where the irregular-satellite and dust-particle orbits are taken to precess around the same axis (normal to Saturn's orbital plane):

\begin{equation}\label{Tcol}
T_{col} \sim \pi \Bigg( sin^2 i_{p} + sin^2 i_{I} \Bigg)^{\frac{1}{2}} \Bigg( \frac{a_{I}}{R_{I}} \Bigg)^2 \Bigg( \frac{U_r}{U} \Bigg) T_{orb}, 
\end{equation}
where the $p$ subscript refers to the dust particle, $i_{p}$ and $i_{I}$ are inclinations measured relative to Saturn's orbital plane, $R_{I}$ is Iapetus' radius and $a_{I}$ its semimajor axis; $U$ is the relative speed between the two objects, and $U_r$ is the radial component of the relative velocity; finally, $T_{orb}$ is the dust particle's orbital period \citep{Opik51, Hamilton94}.  Iapetus' orbit does not quite precess around the normal vector to Saturn's orbital plane, but rather around an axis (the normal to its local Laplace plane, see Fig. \ref{Laplace}) $\sim 11^{\circ}$ away \citep{Ward81}; however, for our rough estimate this can be ignored.

Qualitatively, one should expect $T_{col}$ to decrease as $i_p$ approaches $0^{\circ}$ (or $180^{\circ}$).  As $i_p$ approaches coplanarity, the phase space that the particle must explore before ``finding" Iapetus is reduced.  One should also expect prograde particles to have a substantially decreased chance of striking Iapetus compared to retrograde particles, as retrograde particles have a much larger relative velocity $U$ than prograde particles.  This occurs because for larger relative velocities, when one particle is passing through the node where collisions are possible, the other particle can initially be at a wider range of positions in its orbit and still reach the node "in time."

For prograde particles with low inclinations, the azimuthal component of $U$ is mostly subtracted out so $U_r/U$ is roughly 1, and nearly independent of the particle eccentricities \citep{Hamilton94}.  For retrograde particles, on the other hand, the azimuthal component of $U$ dominates so $U_r/U$ will be small and will increase with orbital eccentricity, which determines the departure of the dust orbit from being purely azimuthal.   One should therefore expect that for retrograde particles, those with inclinations closest to coplanarity and with low eccentricities will have the shortest collision timescale and the highest collision probability (see Fig. \ref{irr}).  

We can also investigate this more quantitatively.  The ratio $U_r$/$U$ can be obtained from Eqs. \eqref{vI} and (15), yielding

\begin{equation}\label{Ur}
\frac{U_r}{U} = \Bigg[ 1 + \frac{\alpha^2(1 + \alpha^2 - 2\alpha\:cos i')}{e^2 - (\alpha^2 - 1)^2}\Bigg]^{-\frac{1}{2}},
\end{equation}
where $i'$ is the mutual inclination between the particle's and Iapetus' orbits, $e$ is the particle's eccentricity, and 
\begin{equation}\label{alphasquared}
\alpha^2 = \frac{a_p(1 - e^2)}{a_I}.
\end{equation}
We note that while particle orbits will perform small oscillations in their inclinations around their parent body's inclination, particle eccentricities will be substantially larger than those of the parent bodies due to radiation pressure.  In this section we therefore take $i_p \approx i_{satellite}$ and select characteristic eccentricities from our numerical integrations.  It is nevertheless generally true that source satellites with more eccentric orbits will yield particle orbits with higher eccentricities.

Since particles can only collide with Iapetus when their orbit's pericenter is smaller than $a_I$ and their apocenter is greater than $a_I, \alpha^2$ ranges between $1 - e$ and $1 + e$.  Furthermore, from Eq. \ref{cosiprime}, $i'$ varies between $i_p - i_I$ and $i_p + i_I$.  Taking $i' \approx i_p$ and $\alpha \approx 1$ as characteristic values, and expanding $cos i'$ to leading order,
\begin{equation}\label{UrSimple}
\frac{U_r}{U} = \Bigg[ 1 + \Bigg(\frac{i_p}{e}\Bigg)^2\Bigg]^{-\frac{1}{2}}.
\end{equation}
Since $sin^2 i_I \ll sin^2 i_p$ for Saturn's prograde irregular satellites ($35^{\circ} < i_p < 50^{\circ}$), we can approximate the first term in parentheses in Eq. \eqref{Tcol} as simply $sin i_p \approx i_p$.  Therefore,
\begin{equation}\label{prograde}
T_{col} \sim i_p\Bigg[ 1 + \Bigg(\frac{i_p}{e}\Bigg)^2\Bigg]^{-\frac{1}{2}}\: (prograde).
\end{equation}
Over the range of prograde inclinations and the range of characteristic particle eccentricities found in our numerical integrations ($0.3 \lesssim e \lesssim 0.6$), $T_{col}$ varies by a factor between $\sim 0.3-0.45$.  This approximation agrees with the values from Eq. (prograde) to within $25\%$ over the range of characteristic values for $e, i_p$ and $a_p$ for dust from the prograde satellites, and matches the small spread ($\lesssim 50\%$) in collision probability for the prograde satellites (plus signs) in Fig. \ref{irr}.  

For the retrograde satellites, again taking $i' \approx i_p$ and $\alpha \approx 1$, and noting that $\pi - i_p$ is small so that $cosi_p \approx -1 + (\pi - i_p)^2/2$,
\begin{equation}\label{TcolRet}
T_{col} \sim (\pi - i_p)\Bigg[1 + \frac{4}{e^2}(1 - \frac{(\pi - i_p)^2}{4})\Bigg]^{-\frac{1}{2}}.
\end{equation}
Thus, since $(\pi - i_p)^2/4 \ll 1$ for the retrograde satellites and $4/e^2 \gg 1$ for the range of characteristic eccentricities ($\gtrsim 0.3$),
\begin{equation}\label{retrograde}
T_{col} \sim (\pi - i_p)\: \frac{e}{2} \: (retrograde).
\end{equation}
This agrees with the values from Eq. \eqref{retrograde} to within $20\%$ over the range of characteristic values of $e, i_p$ and $a_p$ for dust from the retrograde satellites.  It also shows why the retrogrades have much higher collision probabilities.  For the range of dust inclinations and eccentricities, the prograde to retrograde ratio of $T_{col}$ using Eqs. \eqref{prograde} and \eqref{retrograde} is always greater than unity and $\lesssim 25$.  

\begin{figure}[!ht]
\includegraphics[width=9cm]{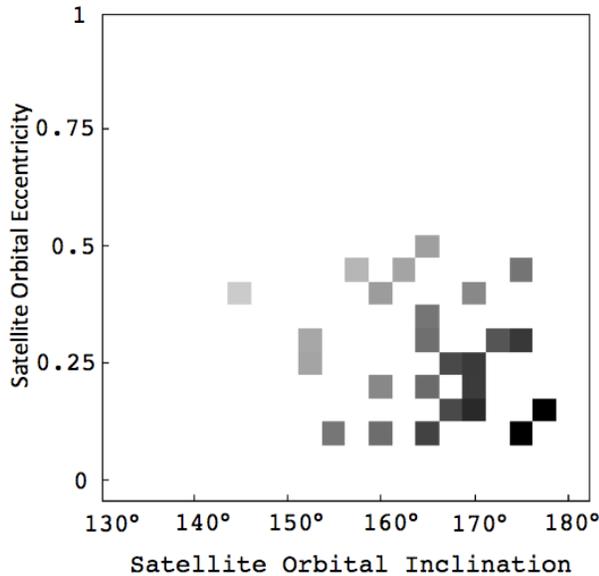}
\caption{\label{2dirr} Numerically computed probabilities for 10-$\mu$m dust particles striking Iapetus as a function of parent (retrograde) satellite orbital inclination and eccentricity.  Probabilities range from darkest (Suttungr $=$ 0.9 and S$/2007\_$S2 $=$ 0.89) to lightest (Narvi $=$ 0.22).  Collision probability increases as the inclination approaches coplanarity ($180^{\circ}$) and as the eccentricity decreases.  }
\end{figure}

Eq. \eqref{retrograde} means that low-eccentricity moons (those that yield lower eccentricity particles) with inclinations close to $180^{\circ}$ will have the shortest collision timescales, and therefore the largest collision probabilities with Iapetus.  Fig. \ref{2dirr} shows the same probabilities as Fig. \ref{irr}, but in two dimensions so as to separate the dependence on inclination and eccentricity.  Darker squares represent larger collision probabilities.  Thus one can see following rows of constant eccentricity that the probability increases as the inclination approaches $180^{\circ}$, whereas following columns of constant inclination, the probability decreases with increasing eccentricity.  \cite{Tosi10} find similar trends using a different method of evaluating collision probabilities.  The two low-eccentricity, high-inclination moons on the bottom right of the plot with the highest probabilities are Suttungr and S$/$2007$\_$S2.

\subsection{Dust Generation Efficiencies}

While the previous section addressed the likelihood of particles from different irregulars striking Iapetus once they are ejected, one must still determine the relative dust yield from the various satellites to infer the dominant sources of dust for Iapetus.  Dust will be generated both in collisional break-up between the outer irregular satellites \citep{Bottke10} and in micrometeoroid bombardment from outside the Saturnian system \citep{Burns99}.  In both cases, the effectiveness of a satellite as a dust source is determined by the competition between a larger satellite radius raising the collision cross section and a larger satellite mass increasing the escape velocity (thus inhibiting dust from leaving the satellite).  

\begin{figure}[!ht]
\includegraphics[width=9cm]{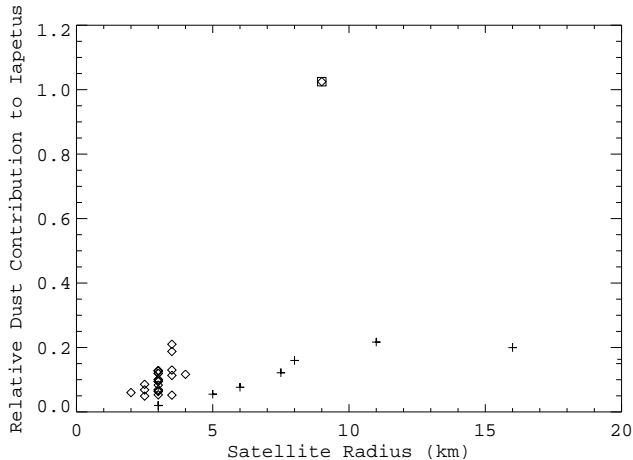}
\caption{\label{rsquared} Dust supplied to Iapetus from each of the irregular satellites relative to the contribution from Phoebe, plotted vs. satellite radius.  The relative contribution is calculated as the product of the collision probability for the particular satellite (Sec. 4) and the radius-dependent terms in the dust-generation efficiency factor of \cite{Burns99}, assuming an optimum satellite size of 10 km.  Prograde satellites are represented by plus signs, and retrograde moons by open diamonds.  Ymir is plotted as a boxed diamond ($R_i = $9 km) and has the largest contribution.}
\end{figure}

\cite{Burns99} investigate this relationship in the Jovian system.  For small moons, where gravity is not important, the rate at which mass is supplied to the ring by a satellite of radius $R_i$ is simply proportional to ${R_i}^2$; however, beyond an optimum satellite size $R_{opt}$ that depends on regolith properties, the dust production rate becomes almost flat, decreasing as ${R_i}^{-1/4}$.  \cite{Burns99} estimate that this optimum size should be about $5-10$ km in the Jovian system.  Assuming similar results for the Saturn system, this implies that Phoebe ($R\sim100$ km) should produce no more (in fact slightly less) dust than any $\sim 5-10$ km irregular satellite.  Fig. \ref{rsquared} shows the impact probabilities calculated in the previous section for each irregular satellite weighted by the $R_i$ dependent terms in the dust-generation efficiency factor of \cite{Burns99}, assuming an optimum satellite size $R_{opt}$ of 10 km.  

Fig. \ref{rsquared} shows that Ymir ($R_i = $ 9 km) should be roughly as important a contributor of dust to Iapetus as Phoebe ($R_i = $ 107 km), though the summed contribution from the remaining moons is greater than that of either Ymir or Phoebe.  This might help lessen the contradiction that the spectrum of Cassini Regio does not seem to match that of Phoebe \citep{Buratti05, Tosi10}.  

But if Phoebe is not the dominant source of dust in the outer Saturnian system, why then is the only prominent dust ring generated by the irregular satellites associated with Phoebe?  Satellites should generate dust rings of height $2asin\:i$, so one might expect to see a nested series of rings of differing heights.  This is analogous to the dust bands observed in the zodiacal cloud, where the dimensions of the bands give away the orbital elements of the object that produced them \citep{Dermott84}. 

Perhaps the fact that Phoebe's orbit has the smallest semimajor axis and lowest inclination among the irregular satellites squeezes its modest share of dust into a more compact volume, yielding a higher optical depth than other satellite rings.  The line-of-sight optical depth of a ring generated by satellite $j$, $\tau_j \sim n_j\sigma_j L_j$ where $n_j$ is the number density, $\sigma_j$ is the average particle cross-section and $L_j$ is the distance along the line of sight.   Focusing only on the parameters involving the satellites (as opposed to dust properties), and taking $L_j \sim a_j$, $\tau_j \sim M_j a_j/V_j$, where $V_j$ is the volume of the ring associated with satellite $j$ and $M_j$ is the mass of dust within it.  $V_j$ should be proportional to $a_j^3sin i_j$, yielding 

\begin{equation}\label{tauj}
\tau_j \sim \frac{M_j}{a_j^2 sin\:i_j}
\end{equation}

\begin{figure}[!ht]
\includegraphics[width=9cm]{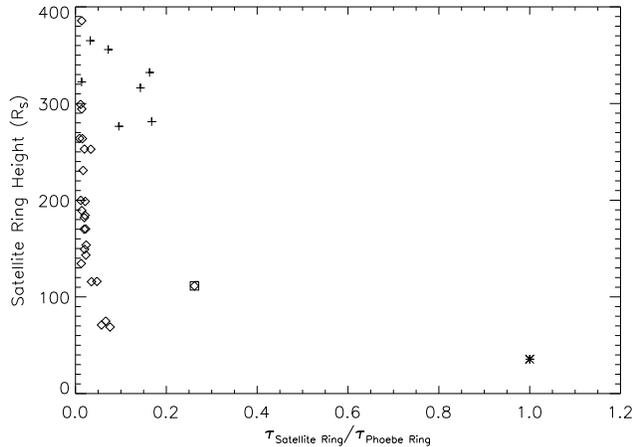}
\caption{\label{tauvsheight}  Estimated line-of-sight optical depth of rings created by the irregular satellites relative to the optical depth of the Phoebe ring, plotted vs. the height of the ring that the satellite would produce in Saturn radii.  Phoebe (asterisk) generates the ring of highest optical depth with a thickness of $\sim$ 40 $R_S$, followed by Ymir (boxed diamond), which should produce a $\sim 110 R_S$-tall ring. Plus signs denote regular satellites, open diamonds irregular moons.}
\end{figure}

The mass $M_j$ carries the same weighting factor used in Fig. \ref{rsquared}.  The optical depth therefore carries an additional factor of ${(a_j^2 sin\:i_j)}^{-1}$.  Fig. \ref{tauvsheight} plots the weighting factor in Eq. \eqref{tauj} relative to that for Phoebe vs. the expected ring height that would be produced by the particular moon ($2asin\:i$) in Saturn radii.

The simple estimates illustrated in Figs. \ref{rsquared} and \ref{tauvsheight} suggest that irregular satellites other than Phoebe could contribute substantial amounts of dust to Iapetus, while the Phoebe ring (due to its compactness) would be the most prominent dust ring generated by the irregulars.  Perhaps in observations of the Phoebe ring, the flux from the various much taller, low-optical depth rings has been interpreted as part of the background.  One might still, however, expect to be able to identify a dust ring associated with Ymir; it would be $\sim 3$ times taller, and have $\sim 1/4$ the line-of-sight optical depth of Phoebe's structure.

The fact that no other dust rings have yet been detected could be used to argue that other factors enhance Phoebe's dust production in the outer Saturn system.  One possibility is that Phoebe's position as the innermost irregular satellite increases its collision frequency with other irregulars.  Numerical studies \citep{Nesvorny03} suggest that Phoebe alone among the Saturnian irregulars likely suffered collisions with several now absent irregulars; while below the detection threshold of today's telescopes, the ejecta from these events would be excellent suppliers of debris.  As such, there might be an increased amount of unseen collisional debris ($\lesssim 1$ km) sharing Phoebe's orbital elements, which, for a steep enough size distribution, could contribute significantly to the Phoebe ring.  A more certain assessment will have to await further observations of the Phoebe ring and searches for separate dust bands.  Our studies assume that Phoebe is the dominant dust source in the outer Saturn system.  Should evidence to the contrary arise, further work would be required to assess the relative contributions to Iapetus, Hyperion and Titan.

\section{Implications Beyond Iapetus}

\subsection{Iapetus as a Tracer of the Initial Dust Mass at Saturn}

\cite{Bottke10} argue that most of their estimated $10^{20}$ kg of collisionally generated dust in the outer Saturn system was created within a few hundred Myr of the capture of the irregular satellites.  In this case, the whole range of particle sizes we considered should have had time to reach Iapetus, corresponding to the $\gamma \approx 0$ case discussed in Sec. 3.2.  In this circumstance, since particles roughly larger than $10 \mu$m are almost certain to strike Iapetus, we should expect all the dust mass in sizes $\gtrsim 10 \mu$m to be part of Cassini Regio.

Unfortunately, as \cite{Bottke10} point out, $10^{20}$ kg would generate a dark layer on Iapetus that is kilometers thick.  Taking the depth on the leading side to fall off as $cos\: \psi$, we can estimate the volume of dust on Iapetus as  
\begin{equation}\label{VIap}
V_{Iap} \sim \pi R^2 \: d_{apex} \approx 850 \Bigg(\frac{d_{apex}}{50 cm}\Bigg) km^3,
\end{equation}
where $d_{apex}$ is the peak depth of dust at the apex (not including any lag deposit from sublimation).  Assuming 100$\%$ transfer efficiency (all particles $\gtrsim 10 \mu$m), this would imply an initial dust mass $\sim 10^{15}$ kg.

The transfer efficiency could be reduced with a sufficiently steep size distribution (one that would cause essentially all of the dust mass to be in sizes $\lesssim 5 \mu$m).  These small particles would then be quickly eliminated from the system through radiation pressure and avoid Iapetus; unfortunately, the requisite power-law index is implausibly steep ($\gtrsim 5.5$).  

The two results might also be reconciled if Iapetus achieved synchronous rotation much later ($\gtrsim 1 Gyr$ after formation).  In this case, the dark material would have been localized to the leading side only after the bulk of the influx had occurred.   Such a scenario could help explain why even the bright hemisphere of Iapetus is darker than the surfaces of the other large icy satellites; however, it also poses the problem of why the masses of dust that would have blanketed Iapetus at all longitudes prior to synchronous rotation did not cause blackening everywhere when thermal migration \citep{Spencer10} kicked in.  

It seems for now that, unless the depth estimate derived from radar measurements \citep{Ostro06} is grossly in error, the amount of dark material on Iapetus implicates an initial dust mass in the outer Saturnian system about 5 orders of magnitude lower than that of \cite{Bottke10}.  While the presence of ammonia could make a thicker layer of dark material appear shallow \citep{Ostro06}, the discovery of small bright-floored craters close to the boundary of Cassini Regio support the idea of a thin dark deposit \citep{Denk10}.

\subsection{Hyperion}

During Cassini's close fly-by, \cite{Cruikshank07} and \cite{Thomas07} found Hyperion to be segregated into a low-albedo unit mostly filling the bottoms of cup-like craters and a more widespread high-albedo unit.  Furthermore, the spectra of the dark material show similarities to the material making up Cassini Regio, suggesting a common source \citep{Cruikshank07, Buratti05}.  

\cite{Burns96}, in their dynamical study of the fate of Phoebe dust, had already argued that Hyperion should receive a significant fraction of Phoebe dust grains.  In their calculations, however, they considered collisions with Iapetus, Hyperion and Titan sequentially, when in fact---because of radiation-pressure-induced orbital eccentricities---dust grains can reach all three moons nearly simultaneously.  As a result, Hyperion will receive a much-reduced fraction of the grains ($\sim 0.004$ for both 5- and 10-$\mu$m grains vs. 0.18 according to \cite{Burns96}).  Furthermore, no grains $\gtrsim 10 \mu$m reach Hyperion.  The conclusion, however, is the same---Hyperion's surface layers should contain some dark material from Phoebe and the other irregular satellites.  Furthermore, since Hyperion is chaotically rotating rather than tidally locked, an isotropic distribution of dust is expected \citep{Burns96}.

We estimate the volume of material striking Hyperion relative to the volume hitting Iapetus as

\begin{equation}\label{VHyp}
\frac{V_{Hyp}}{V_{Iap}} \sim \frac{\int_{5 \mu m}^{10 \mu m} r^{3 - (\beta + \gamma)}P_H(r) dr}{\int_{10 \mu m}^{1 cm} r^{3 - (\beta + \gamma)} dr},
\end{equation}
where $P_H(r)$ is the probability for a particle of size $r$ striking Hyperion (derived from our numerical simulations discussed in Sec. 2), $V_{Iap}$ is given in Eq. \eqref{VIap}, and we have approximated the probability of particles striking Iapetus as a step function at $r = 10 \mu$m. 

\begin{figure}[!ht]
\includegraphics[width=9cm]{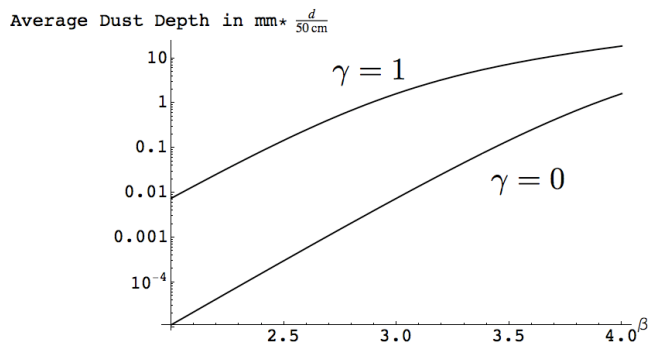}
\caption{\label{hyp} Average dust depth on Hyperion (in mm) vs power-law index $\beta$, for the limiting cases of $\gamma = 0$ (bottom curve) and $1$ (top curve).}
\end{figure}

Calculated dust depths on Hyperion for $2 < \beta < 4$ in the limiting cases of $\gamma = 0$ and $1$ (cf. Sec. 3.2) are given in Fig. \ref{hyp} assuming isotropic coating and a spherical target of radius 135 km.

These extremely shallow depths of $\lesssim 1$ cm render it plausible that Hyperion might not be uniformly covered in dust, though the mechanism for segregating dark material to the bottoms of Hyperion's ubiquitous sharp-edged craters as is observed remains unclear \citep{Cruikshank07}.  Constraints on the particle size distribution in the Phoebe ring from future observations may narrow estimates of the material delivered to Hyperion.

\subsection{Titan}

As discussed in Sec. 2.4, Titan's much larger cross-section causes it to efficiently sweep up almost everything that crosses its orbit.  The slow inward migration of dust particles $\gtrsim 10 \mu$m gives Iapetus enough time to collect most of them before they become Titan-crossing; our numerical simulations of Sec. 2, however, indicate that $\sim 70\%$ of $5\mu$m particles strike Titan (and that smaller particles are so affected by radiation pressure that they either strike Saturn in the first half-Saturn year or escape the system).  As in the case with Iapetus, particles will strike Titan on its leading side; however, its atmosphere will distribute material around the entire moon and will fragment particles upon entry.

Using Cassini's measurements, \cite{Porco05} report a detached haze layer at 500-km altitude on Titan.  Since the sedimentation time in this layer is short, some process must continually replenish particles.  Two hypotheses have been proposed \citep{Tomasko09}:  the layer either represents a condensation region at a local temperature minimum \citep{Liang07} or occurs where aerosols produced at higher altitudes settle\citep{Lavvas09}.  

\begin{figure}[!ht]
\includegraphics[width=9cm]{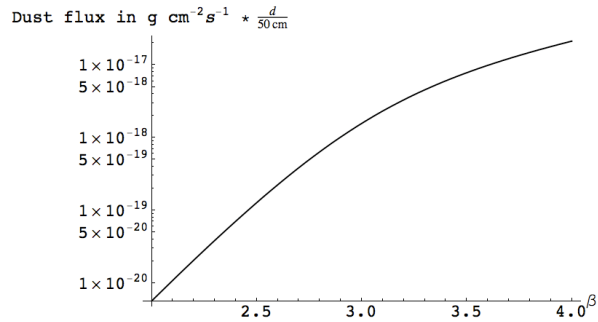}
\caption{\label{titan} Dust flux into the Titan atmosphere at 500 km altitude (in $g\: cm^{-2} \: s^{-1}$) vs. power-law index $\beta$ for the limiting case of $\gamma = 1$.}
\end{figure}

While this haze layer is likely due to the above-mentioned atmospheric effects, we can explore whether exogenous dust-infall might also be a significant contributor.  We therefore derive the volume of particles striking Titan as we did for Hyperion (Eq. \ref{VHyp}) assuming a present rate of dust production that is constant in time ($\gamma = 1$).  Fig. \ref{titan} shows the results, expressed as a mass flux computed $\sim$ 500 km above the Titan surface (i.e., $R$ = 3100 km).

The calculated mass flux falls well short of the estimated 2.7-4.6 \e{-14} $g\:cm^{-2}\:s^{-1}$ \citep{Lavvas09} required to replenish particles.  Fortunately, the mechanisms listed above seem sufficient for explaining the haze layer (R. A. West, personal communication, 2010).  

\section{Conclusion}

Our results show that out of the dust particles collisionally generated at Phoebe that are long-lived (grains $\gtrsim 5 \mu$m), most larger than $\sim 10 \mu$m will strike Iapetus due to modifications of their orbits by perturbations from the Sun.  The latter include Poynting-Robertson drag, solar radiation pressure and the Sun's tidal gravity in the Saturn system.  

Our computed dust coverage on the Iapetus surface matches up well with the newly discovered color dichotomy on Iapetus that extends up to the poles \citep{Denk10}.  The calculated distribution also traces the shape of Cassini Regio well in the longitudinal direction, but realistic thermal modeling is required to explain both the bright poles and the sharp boundaries between bright and dark material.  

Our orbital calculations for $10 \mu$m particles show that dust launched from other retrograde outer irregular satellites can have comparable likelihoods of striking Iapetus to those of dust launched from Phoebe.  We argue this can contribute to the differing spectra between Phoebe and Iapetus; however, the question of how much dust was generated by Phoebe relative to the other irregular satellites is still unclear.

By tracking the dust that strikes Hyperion, we find that just a veneer should have been laid down on its surface ($\lesssim 1$ cm on average).  This picture may be consistent with the observation that the surface is not uniformly coated.  \cite{Cruikshank07} and \cite{Thomas07} find the dark material to be predominantly at the bottoms of Hyperion's ubiquitous cup-like craters.  As opposed to Iapetus, which is tidally locked, Hyperion rotates chaotically, which can explain the presence of dark material throughout the surface.

We determine that effectively all long-lived dust particles that avoid Iapetus (i.e., a fraction of those between $\sim 5$ and $10 \mu$m) are swept up by Titan.  While these constitute a considerable mass flux into the Titan atmosphere, they are insufficient to account for the satellite's detached haze layer.

\section{Acknowledgements}
We would like to thank Philip D. Nicholson, Matthew S. Tiscareno, Rebecca A. Harbison, Anthony J. Milano and Robert A. West for insightful comments and discussions, as well as Tilmann Denk and Juergen Schmidt for careful and constructive reviews.  We acknowledge funding support from the Cassini project and from NASA's Planetary Geology and Geophysics Program.

\end{document}